\documentclass[sigconf, nonacm]{acmart}
\usepackage[ruled,vlined,linesnumbered]{algorithm2e}
\usepackage{enumitem}
\usepackage{multirow}
\usepackage{mdframed}
\usepackage{minted}
\usepackage{xcolor}
\usepackage{tikz}
\usepackage{subcaption}
\usepackage{booktabs}
\usepackage{caption}
\usepackage{algpseudocode}

\usepackage{circledtext} % for cicled numbers
\circledtextset{resize=real}
\usepackage{pifont} % for cicled numbers inverted
\usetikzlibrary{arrows.meta, positioning}
\usepackage[most]{tcolorbox}

\usepackage{tcolorbox}
\tcbuselibrary{breakable}

\newcommand\vldbdoi{XX.XX/XXX.XX}
\newcommand\vldbpages{XXX-XXX}
\newcommand\vldbvolume{14}
\newcommand\vldbissue{1}
\newcommand\vldbyear{2020}
\newcommand\vldbauthors{\authors}
\newcommand\vldbtitle{\shorttitle} 
\newcommand\vldbavailabilityurl{URL_TO_YOUR_ARTIFACTS}
\newcommand\vldbpagestyle{plain}

\newcommand{\joey}[1]{\textcolor{orange}{\textbf{\footnotesize [joey: #1]}}}
\usepackage{xspace}

\newcommand{\sys}{\textsc{Tk-Boost}\xspace}

\DeclareMathOperator*{\concat}{\texttt{concat}}

\usepackage{minted}

\setminted{
  bgcolor=gray!4,
  fontsize=\small,
  breaklines=true,
  escapeinside=||   % <-- important
}

\newif\ifcomments
\commentstrue
\ifcomments
    \providecommand{\alvin}[1]{{\protect\color{purple}{\bf [alvin: #1]}}}
\else
    \providecommand{\alvin}[1]{}
\fi

\begin{document}
\title{Arming Data Agents with Tribal Knowledge}
\author{Shubham Agarwal$^{\dagger}$, Asim Biswal$^{\dagger}$, Sepanta Zeighami$^{\dagger}$ \\Alvin Cheung, Joseph Gonzalez, Aditya G. Parameswaran}
\affiliation{%
UC Berkeley\\
\{\url{shubham3, abiswal, zeighami, akcheung, jegonzal, adityagp}\} \url{@ berkeley.edu}
\thanks{$^\dagger$Equal contribution in alphabetical order}
}

%%
%% The "author" command and its associated commands are used to define the authors and their affiliations.
% \affiliation{%
%   \state{Ireland}

% \orcid{0000-0002-1825-0097}
% \affiliation{%
%   \country{Iceland}

% \orcid{0000-0001-5109-3700}
% \affiliation{%
% \email{vb@rocquencourt.com}

% \author{J\"org von \"Arbach}
% \affiliation{%
%   \institution{University of T\"ubingen}
%   \city{T\"ubingen}
%   \country{Germany}
% }
% \email{jaerbach@uni-tuebingen.edu}
% \email{myprivate@email.com}
% \email{second@affiliation.mail}

% \author{Wang Xiu Ying}
% \author{Zhe Zuo}
% \affiliation{%
%   \institution{East China Normal University}
%   \city{Shanghai}
%   \country{China}
% }
% \email{firstname.lastname@ecnu.edu.cn}

% \author{Donald Fauntleroy Duck}
% \affiliation{%
%   \institution{Scientific Writing Academy}
%   \city{Duckburg}
%   \country{Calisota}
% }
% \affiliation{%
%   \institution{Donald's Second Affiliation}
%   \city{City}
%   \country{country}
% }
% \email{donald@swa.edu}

\setlength{\abovedisplayskip}{3pt}
\setlength{\belowdisplayskip}{3pt}
\setlength{\abovedisplayshortskip}{3pt}
\setlength{\belowdisplayshortskip}{3pt}
\setlength{\textfloatsep}{3pt}
\setlength{\floatsep}{3pt}
\setlength{\intextsep}{3pt}
\setlength{\dbltextfloatsep}{3pt}
\setlength{\dblfloatsep}{3pt}
\setlength{\abovecaptionskip}{3pt}
\setlength{\belowcaptionskip}{3pt}

% Set list appearance
\setlist[itemize]{leftmargin=*, topsep=0.5em}

%\titlespacing*{\section}{0pt}{0.8ex}{0.5ex}
%\titlespacing*{\subsection}{0pt}{0.9ex}{0.7ex}
%\titlespacing*{\subsubsection}{0pt}{0.8ex}{0.7ex}

\begin{abstract}
Natural language to SQL (NL2SQL) translation enables non-expert users to query relational databases through natural language. Recently, NL2SQL agents, powered by the reasoning capabilities of Large Language Models (LLMs), have significantly advanced NL2SQL translation. Nonetheless, NL2SQL agents still make mistakes when faced with large-scale real-world databases because they lack knowledge of how to correctly leverage the underlying data (e.g., knowledge about the intent of each column) and form \textit{misconceptions} about the data when querying it, leading to errors. Prior work has studied generating facts about the database to provide more context to NL2SQL agents, but such approaches simply restate database contents without addressing the agent's misconceptions. In this paper, we propose \sys, a bolt-on framework for augmenting any NL2SQL agent with \textit{tribal knowledge}: knowledge that corrects the agent's misconceptions in querying the database accumulated through experience using the database. To accumulate experience, \sys 
first asks the NL2SQL agent to answer a few queries on the database, identifies the agent's misconceptions by analyzing its mistakes on the database, and generates tribal knowledge to address them. To enable accurate retrieval, \sys indexes this knowledge with \textit{applicability conditions} that specify the query features for which the knowledge is useful. When answering new queries, \sys uses this knowledge to provide feedback to the NL2SQL agent, resolving the agent's misconceptions during SQL generation, and thus improving the agent's accuracy. Extensive experiments across the BIRD and Spider 2.0 benchmarks with various NL2SQL agents shows \textbf{\sys improves NL2SQL agents accuracy by up to 16.9\% on Spider 2.0 and 13.7\% on BIRD}. 
\end{abstract}

\maketitle
\pagestyle{\vldbpagestyle}
\if 0
\begingroup\small\noindent\raggedright\textbf{PVLDB Reference Format:}\\
\vldbauthors. \vldbtitle. PVLDB, \vldbvolume(\vldbissue): \vldbpages, \vldbyear.\\
\href{https://doi.org/\vldbdoi}{doi:\vldbdoi}
\endgroup
\begingroup
\renewcommand\thefootnote{}\footnote{\noindent
This work is licensed under the Creative Commons BY-NC-ND 4.0 International License. Visit \url{https://creativecommons.org/licenses/by-nc-nd/4.0/} to view a copy of this license. For any use beyond those covered by this license, obtain permission by emailing \href{mailto:info@vldb.org}{info@vldb.org}. Copyright is held by the owner/author(s). Publication rights licensed to the VLDB Endowment. \\
\raggedright Proceedings of the VLDB Endowment, Vol. \vldbvolume, No. \vldbissue\ %
ISSN 2150-8097. \\
\href{https://doi.org/\vldbdoi}{doi:\vldbdoi} \\
}\addtocounter{footnote}{-1}\endgroup

\ifdefempty{\vldbavailabilityurl}{}{
\vspace{.3cm}
\begingroup\small\noindent\raggedright\textbf{PVLDB Artifact Availability:}\\
The source code, data, and/or other artifacts have been made available at \url{\vldbavailabilityurl}.
\endgroup
}
\fi

%\textcolor{red}{---------Aditya is taking a pass below---------}
\section{Introduction}\label{sec:intro}
Natural language (NL) interfaces to databases enable users without SQL expertise to query data in relational databases using natural language. % without writing SQL, democratizing data access for users without SQL expertise. To enable an NL interface, the database translates a given query in NL to its equivalent SQL statement. 
In recent years, the proficiency of Large Language Models (LLMs) in writing SQL, coupled with tool-calling capabilities~\cite{hong2025next, yao2023react} has led to significant breakthroughs in accurate NL2SQL translation~\cite{11095853, deng2025reforce, li2024dawn, schmidt2025sqlstorm}. These \emph{NL2SQL agents} accept NL queries and return SQL, and 
vary across prompting strategies, provided tools (e.g., database or file system access), and workflows they are embedded in (e.g., existence of validation loops or consistency mechanisms~\cite{li2024dawn, ren2024purple, sun2024r}).  
%NL2SQL systems involve one more NL2SQL agents to perform this translation. Differences across systems arise from how the agents are embedded, , or which prompting strategies or tools are given to the agents .

%, and a large body of work has studied how to use LLMs for accurate translation. %give LLMs access to the database to explore the schema or execute partial queries. They 
% We refer to an LLM-powered function that generate a SQL query given an NL query as an \textit{NL2SQL agent}. 
% NL2SQL systems use NL2SQL agents to perform NL2SQL translation, with differences across systems in how the agents are embedded within them, e.g., whether they use validation loops or consistency mechanisms~\cite{li2024dawn, ren2024purple} as well as whether any prompting strategies and tools given to the agents (e.g., database access or file system access provided or not).
%NL2SQL agents are the backbone of current state-of-the-art NL2SQL methods, with their differences primarily in 

\textbf{Limitations of NL2SQL Agents}. Despite significant improvements in recent years, agents still make mistakes in large real-world databases, with many columns, tables, and rows. 
This is due to gaps in the agent's knowledge about databases not seen during training set---such as its contents, how it was created and its intended use. Such knowledge gaps, especially about subtle data idiosyncrasies, 
cause logical errors in translation~\cite{liu2025nl2sql, chung2025long}, where agents use incorrect data sources (tables or columns) or perform logically incorrect operations on them. We refer to these recurring logical errors in performing queries caused by the agent's knowledge gaps as \textit{misconceptions}. 
For example, multiple tables or columns with similar semantics are common in data lakes, where derived tables with modified columns (e.g., NULLs removed through imputation), are stored alongside the originals, causing misconceptions when deciding which column to use. 
We use such a scenario as our running example: a table containing product sales information with columns \texttt{name} and \texttt{brand}, with \texttt{brand} inconsistently populated (perhaps because it was added later on), while \texttt{name} reliably contains both the product name and brand, with the latter as a prefix (e.g., ``Nike Air Max''). An agent using \texttt{brand} for queries such as ``find average sales by brand'' produces incorrect SQL, with errors recurring for any query that requires product brand information.

Prior work has tried to fix misconceptions by proactively identifying "facts", typically  %To facilitate database usage, prior work has studied automatically generating or refining database metadata 
by asking LLMs to generate them about the database content~\cite{zhang2025autoddg, balaka2025pneuma, luoma2025snails, zhang2023nameguess, baek2025}.
%A simple solution to address such misconceptions is to first automatically generate meta-data---facts about the tables, columns and data values---to help the agents better understand the database. 
%Such approaches  to inspect database schema and values and summarize observations as meta-data. %These methods surface descriptive facts about the database. 
However, doing so in the absence of queries simply restates this content without fixing the misconceptions that arise when answering queries.
%such metadata does not correct the NL2SQL agent's misconceptions when answering queries---it describes database content and terminology without clarifying
%about which data source to use to answer a query or how. %Agent's misconceptions are often nuanced misunderstandings about data semantics and usage that metadata generated without considering agent's misconceptions requires observing what misconceptions occur and then generating metadata for correcting them. %---misconceptions often do not surface during unguided exploration and are thus commonly not fixed based on the facts generated when doing so. %Therefore, these approaches generates redundant facts that do not help correct agent misconceptions. 
In our example, %involving a table with \texttt{product\_name} and \texttt{brand} columns, 
one fact may be that both \texttt{name} and \texttt{brand} columns contain product brands, %, , while \texttt{brand} contains only the product brand. This 
but this information does not help the agent infer that \texttt{name} should be used instead of the \texttt{brand} column. %Moreover, such metadata is even less useful for correcting data-usage misconceptions that stem from misunderstandings of operation-specific behavior, such as determining how to handle null values produced by a join. 

\textbf{Agents with Tribal Knowledge}. We propose \sys, a framework for correcting NL2SQL agents' misconceptions. % using what we call \textit{tribal knowledge}. 
Our key insight is that an NL2SQL agent should accumulate knowledge of how to correctly query a database \textit{through experience}, rather than relying on facts derived from the data---this process is analogous to how data analysts or engineers accumulate knowledge over time through exposure to the database. A simple approach to accumulate experience is to store the agent's interactions with the database (i.e., queries and answers)---or their summaries---and retrieve relevant ones to include in the agent's prompts for future queries via a form of RAG (retrieval augmented generation) using LLM memory stores~\cite{shinn2023reflexion, xu2025mem, li2025memos, li2025aid, talaei2024chess, mem0} such as Mem0~\cite{mem0}. 

However, as we show, directly including past interactions does not correct \emph{misconceptions} as past mistakes end up being repeated. %,  and thus does not improve accuracy for future queries. % (past interactions include query-specific information that can't be generalized to new queries). 
Instead, \sys augments agents with what we call \textit{tribal knowledge}: natural language statements %containing \textit{corrective} and \textit{reusable} information 
that \textit{correct the agent's misconceptions} and can be \textit{reused to correctly answer related future queries} (e.g., stating ``use \texttt{name} column to find product brands instead of \texttt{brand}'' in our running example). %We call such  \textit{corrective} and \textit{reusable} knowledge \textit{tribal knowledge for data agents}, where each piece of knowledge is a set of natural language statements, containing information that correct an agent's misconception and can be reused to answer future queries. % 
To enable an agent to accumulate tribal knowledge through experience, \sys uses a few labeled query examples (i.e., NL queries with ground-truth SQL) and asks the agent to answer those queries on the database. \sys then \textit{discovers tribal knowledge} from its experience and \textit{uses the knowledge} to improve the agent's accuracy for future queries. Operationalizing these two steps is hard as it requires identifying misconceptions, generating reusable knowledge that fixes them, and using it correctly for future queries, as we discuss next. %; all of these steps involve use of error-prone LLMs that lack database knowledge, and naively using them yields limited gains. 

\begin{figure}[t]
    \centering
    % solid black square, 1cm x 1cm
    \vspace{-1em}\includegraphics[width=\columnwidth]{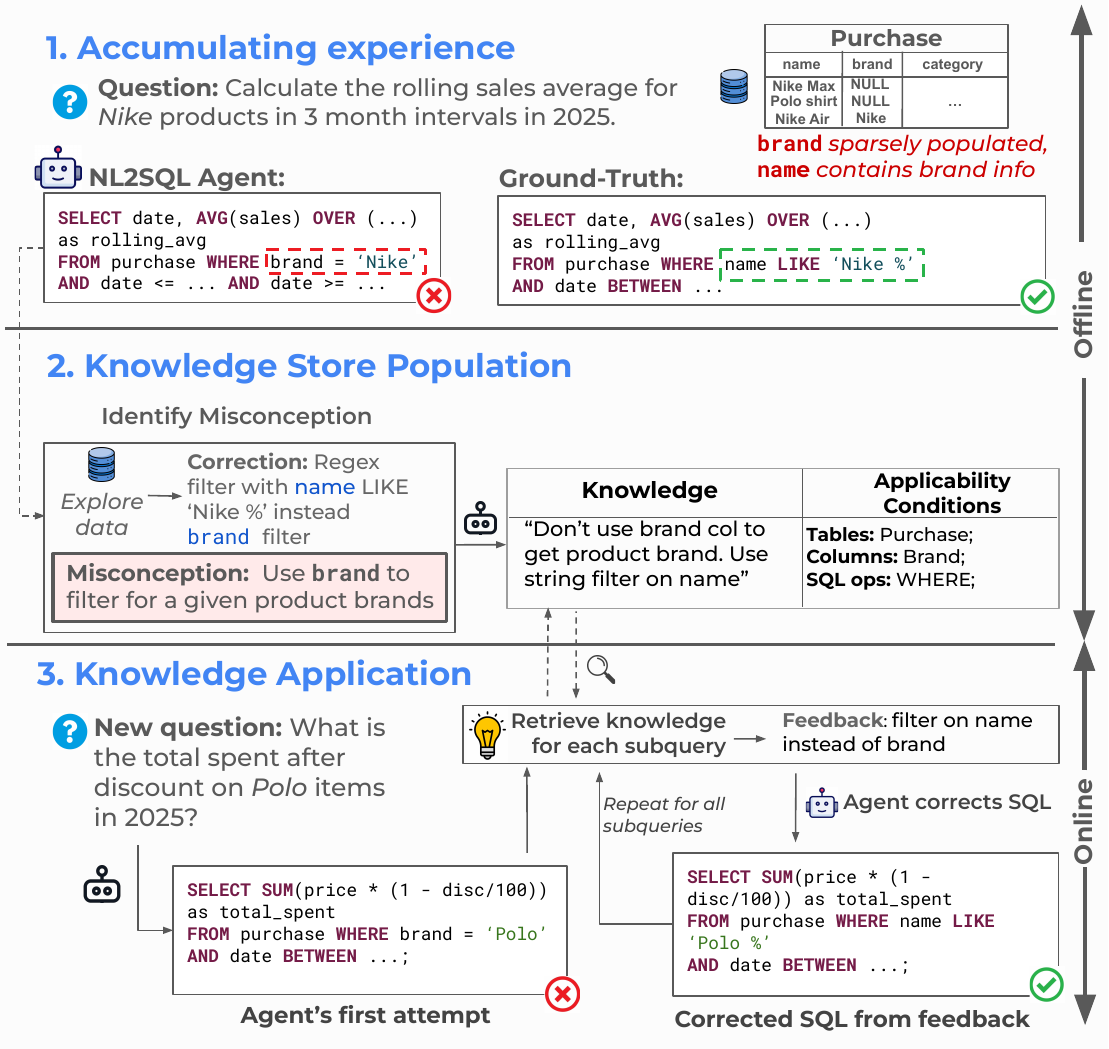}
    \caption{Example workflow of \sys. Past experience is used to find agent misconceptions and generate corrective knowledge. The knowledge is paired with applicability conditions to help retrieval for future queries, and used to correct future agent mistakes.}%Bad vs. good knowledge derived from a training-time NL2SQL error. Basic knowledge contrasting agent answer to gold answer fails to generalize. Knowledge that captures the underlying data-selection misconception helps correct related test-time queries.}
    \label{fig:rule-vs-no-rule-img}
\end{figure}

%Fig.~\ref{fig:rule-vs-no-rule-img} shows an example, where the generated knowledge avoids mentioning column names or specific operators, but contains instructions to avoid using sparsely populated columns in a specific table. % by either specifying when to use a specific table or column for data-selection misconceptions (e.g., specifies a column should be used for certain types of queries) or usage of a SQL operator while abstracting away specific table and column names as much as possible to correct data-usage misconceptions (e.g., specifies how to write mathemtical operations or handle dffierent data types and null values).%so that underlying logical errors must be detected from \textit{complex} SQL queries  

\textbf{Tribal Knowledge Discovery}. \sys identifies agent’s misconceptions through analyzing the agent's \emph{incorrect SQL queries}
and generates tribal knowledge---reusable natural language statements that correct the misconceptions that lead to incorrect queries. 
Accurately identifying misconceptions is challenging because many SQL queries contain complex logic that are difficult to comprehend, while errors depend on data idiosyncrasies so detecting them require an understanding of the underlying data. Simply prompting an LLM to derive knowledge by comparing the agent-generated and ground-truth SQL (similar to summarizing past experience in memory stores~\cite{baek2025knowledge, chhikara2025mem0, li2025memos}) often yields incorrect or trivial statements because the LLM lacks context about database and the complexity of queries cause mistakes. Additionally, future queries are often very different from the small set of past queries the knowledge is generated from. Knowledge statements naively generated with an LLM, even when correct, often contains query-specific knowledge that cannot help address misconceptions for future queries. % and cannot identify the agent's logical errors, especially for complex queries when the two SQL statements appear vastly different syntactically but logical differences are minor. 

To reliably discover tribal knowledge, \sys tries to identify the root cause of the error through iteration on the query and the database. 
% first explores the database to understand relevant data characteristics using which it identifies logical errors in the agent-generated SQL, done by decomposing the process into simpler steps with verifiable guardrails to ensure accurate identification of logical errors with an LLM.  %It does so using a process decomposed into simpler steps with verifiable guardrails to ensure accurate identification of logical errors with an LLM. %, where 
It iteratively makes small data-aware modifications---changing one SQL clause at a time---to the incorrect agent-generated SQL until it produces the correct answer, and detects logical errors along the way.
% from the modifications that made correct logical changes to the SQL. % then evaluates the modifications, and summarizes the rest into a set of NL statements that specify logical errors addressed by the modifications. %---we refer to a sub-component of an NL query as a \textit{sub-task} (e.g., filtering products by brand)---
\sys then generates tribal knowledge statements in NL for each logical error. Each statement explains how to generate a correct SQL not only for the NL query where the logical error was identified, but more broadly to other queries where similar errors may appear, the latter achieved through proactively identifying such queries and generating knowledge statements that correct errors across them. %, enabling reusability of the knowledge statement for future queries. % even if the query itself is very different. % correctly perform the \textit{tasks} where logical errors were encountered, generating general purpose statements that can apply to as many tasks as possible. %These errors occurr for \textit{tasks} affected by the agent's knowledge gap, where we use the term \textit{task} to refer to sub-components of an NL query---thus, NL2SQL translation requires generating and combining SQL statements for tasks within an NL query. 
%, ensuring reusability by abstracting out query specific details. %, thus correcting the misconceptions. 
Fig.~\ref{fig:rule-vs-no-rule-img} illustrates this process. %, a database of product information where the \texttt{brand} column is inconsistently populated while \texttt{name} contains product brand information (Fig.~\ref{fig:rule-vs-no-rule-img} top right corner shows example records). 
The NL2SQL agent accumulates experience by performing a query for which it incorrectly uses the \texttt{brand} column (Step 1). \sys identifies the agent's misconception and generates knowledge that \texttt{name} should be used to obtain product brand information (Step 2). %---the knowledge statement remove. 

\textbf{Tribal Knowledge Usage}. For a new NL query, \sys retrieves relevant knowledge statements and uses it to provide feedback to improve the agent's accuracy. Each statement corrects a specific misconception and must be retrieved and applied for queries where the misconception occurs. Doing so accurately 
is difficult because misconceptions often occur when performing fine-grained sub-tasks required for NL queries (e.g., when joining with a specific table), which are not directly stated in NL queries that are more coarse-grained (e.g., don't mention a join on this table). Thus, naively retrieving knowledge using the NL query---for example, through embedding similarity---yields limited gains because it cannot correctly identify the fine-grained sub-tasks needed for the query. Additionally, when given the relevant knowledge statements, agents often incorrectly use them for sub-tasks they do not apply to, especially for complex queries with many  relevant knowledge statements. Thus, simply injecting all knowledge statements into the prompt often yields limited gains. % because (1) embeddings poorly represent operations needed for a query, causing such retrieval to miss relevant knowledge and (2) especially for complex queries, agents often incorrectly use retrieved knowledge for sub-tasks they do not apply to. %Instead, \sys retrieves knowledge based on the tasks it expects are needed for a query and uses the knowledge to provide targeted feedback for those tasks. To do so, 

Rather than retrieving and applying the knowledge based on the NL query, \sys first asks the NL2SQL agent to write a "first attempt" SQL query, and then retrieves and applies the knowledge to fix logical errors in this SQL. %based on which it provides feedback to the agent to correct the SQL. 
To enable accurate retrieval, \sys indexes the knowledge based on  \textit{applicability conditions} which specify, for each knowledge statement, features of SQL queries (e.g., columns used and SQL clauses present) that may benefit from the knowledge, that is, features of a SQL query wherein the misconception may be present. %help detect what sub-tasks are being performed for an NL query based on the agent's SQL. 
Then, at query time, \sys retrieves knowledge whose applicability conditions include the query features of the agent-generated SQL, uses this knowledge to identify misconceptions in the SQL query, and provides fine-grained feedback to the agent consisting of NL statements on how to correct errors in its SQL. \sys does so one-by-one for each sub-query within the agent-generated SQL, ensuring correct application of the feedback in complex queries. Fig.~\ref{fig:rule-vs-no-rule-img} illustrates this process. In Step 2, the applicability condition of a knowledge statement specifies that it applies to queries accessing the \texttt{brand} and \texttt{name} columns. In Step 3, for a new query, the agent initially generates a SQL query that incorrectly uses \texttt{brand}. After retrieving the relevant knowledge, \sys provides feedback indicating that \texttt{name} should be used instead, leading the agent to produce the correct SQL.

\textbf{Contributions}. \sys is a bolt-on framework to augment any NL2SQL agent with tribal knowledge, providing significant accuracy boosts for any NL2SQL agent by learning how to use a database through experience. %Our evaluation across Bird and Spider 2.0 benchmarks show \sys can be improve accuracy for various NL2SQL agents, providing accuracy boost when using pre-trained models XXX by XXXX, or when using specialized NL2SQL agents XXX by XXX. % By anchoring the knowledge extraction to agent failure modes, \sys addresses the three limitations of prior work: non-reusable knowledge, semantically misaligned retrieval, and unsafe full-query application. 
Specifically, we:
\begin{itemize}
    % \item Present \sys, the first framework that enables any NL2SQL agent to learn how to accurately query a database \textit{through experience};  
    % \item Present \sys, the first framework that enables any NL2SQL agent to learn how to query a database accurately \textit{from experience};
    %show that identifying recurring misconceptions in agent-generated SQL by comparing it to gold SQL provides a reliable signal for recovering database-specific tribal knowledge. We design a \emph{minimal-edit diff} procedure that transforms SQL differences into precise hypotheses about agent misconceptions.
    \item Present \sys, the first framework that enables any NL2SQL agent to accumulate knowledge of how to query a database accurately \textit{from experience};
    \item Introduce the notion of tribal knowledge for data agents to correct their misconceptions;
    \item Show how to reliably discover tribal knowledge from agent's past experience and use it to improve accuracy for future queries; 
    %\item We develop a framework that converts these diffs into \emph{structured rules} with explicit scope over databases, tables, columns, and SQL operators. Rules are retrieved and applied at CTE-level granularity via a validator-driven loop, rather than free-form prompting.
    \item Illustrate through extensive empirical evaluation that \textbf{\sys significantly improves a given NL2SQL agent's accuracy, by up to 16.9\% on Spider 2.0 and 13.7\% on BIRD benchmarks}. These accuracy boosts are up to \textbf{10\% better than memory-based baselines} that provide modest gains of less than 5\%.  
    %\item We demonstrate that leveraging structured tribal knowledge improves NL2SQL accuracy on real-world and benchmark datasets, achieving up to XX\% higher execution accuracy over state-of-the-art LLM-based and memory-augmented baselines, while reducing silent errors from misconceptions by YY\%. LLM-based and memory-augmented baselines~\cite{}.
\end{itemize}
%\textcolor{red}{---------Aditya is taking a pass till here---------}
\section{Background and Problem Statement}
\label{sec:prelims}
% In this section, we formalize the agentic NL2SQL task (Sec~\ref{sec:definition}), define our downstream objective. (Sec~\ref{sec:objective}), and discuss methods for implementing knowledge in knowledge-augmented agents (Sec~\ref{sec:agents}).

% \subsection{Background}
\subsection{NL2SQL Agents}
\label{sec:definition}

\textbf{NL2SQL Translation.}
We study the problem of answering natural language (NL) queries on a relational database. Let $\mathcal{D}$ denote a relational database on which a user issues a natural language query, $q$. Our goal is to generate a SQL query $s$ that is equivalent to $q$. The answer to $q$ is then the result of executing $s$ on $\mathcal{D}$. We denote the result of executing $s$ on $\mathcal{D}$ by $\textsc{ExecSQL}(s, \mathcal{D})$, 
which returns either an output relation or an error string describing an execution failure. 

\textbf{Translation with NL2SQL Agents.}
To perform NL2SQL translation, existing methods use LLMs to translate the given NL query to SQL \cite{gao2023text, 11095853, luo2025natural, gkini2021depth}. We refer to LLM-powered functions that generate SQL for an NL query as an \textit{NL2SQL agent}, and denote it by $\mathcal{A}$. NL2SQL agents are often given access to the database, and query the database to explore the data before generating the final SQL query (they may additionally have access to other tools, e.g., reading text files, which we omit without loss of generality). Specifically, to perform NL2SQL translation, the agent is first provided with the NL query (along with additional prompts or metadata, omitted for simplicity). It then iteratively issues intermediate SQL queries to the database whose results are accumulated into a running \textit{context} and supplied back to the agent at each subsequent iteration. The agent uses this context to generate new SQL queries, eventually generating the final SQL query after which it terminates. % by outputting a termination symbol.  

To formalize this process, denote by $\mathcal{C}$ the space of all possible text strings and by $\mathcal{S}$ the space of all possible SQL queries.  An NL2SQL agent is a function $\mathcal{A} : \mathcal{C} \rightarrow \mathcal{S}\times\{0, 1\}$, implemented using an LLM, that takes a string as input and outputs a SQL query and a Boolean indicator, where the latter denotes whether the output SQL is the final translation result or an intermediary SQL. We refer to the input to the agent, $C \in \mathcal{C}$, as its \textit{input context} or \textit{context} for short. %, which may include the natural-language question, schema information, intermediate SQL drafts, execution results, or external knowledge.
To generate a SQL query for an NL query $q$, the agent is given the query $q$, and proceeds iteratively to generate the SQL query. We denote by $C_t$ the context of the agent at the $t$-th iteration and therefore the \textit{initial context} is $C_0 = q$. At the $t$-th iteration ($t=0$ initially), the agent is given $C_t$ as input and produces a SQL query, $s_t$ along with a Boolean variable, \texttt{is\_final}, that denotes if the query is the final translation result, i.e., $$s_t, \texttt{is\_final} = \mathcal{A}(C_t).$$
$s_t$ is then executed on the database to obtain
$$R_t = \textsc{ExecSQL}(s_t, \mathcal{D}).$$
The output of the SQL query, $R_t$, as well as the query $s_t$ is  concatenated to the context, $C_t$, to create the context for the next iteration,
%, including output relations, execution feedback, and LLM reasoning:
\[
C_{t+1} = \concat(C_t,  s_t,  R_t),
\]
where $\concat$ is a function that concatenates strings. In practice, $C_t$ contains prompts as well as thinking tokens \cite{yao2023react, schick2023toolformer}  (and any other tool use by the agent) which we omit to simplify notation. This process repeats until the agent indicates it has generated the final SQL, $s_n$, at the $n$-th iteration with $n$ denoting the total number of iterations. % which we refer to as translation result. %We refer to each iteration in the above process as a \textit{step}. 
We refer to $C_{n+1}$ as the agent's \textit{execution trace}, which contains all intermediary steps, the final SQL generated and its output. 

% \joey{Minor notational issue.  You use $\mathcal{M}$ for LLMs and then your augmenting mechanism is also $M$.  Maybe just use \texttt{LLM} for LLM?}
\textbf{LLMs and Embedding Models}. In addition to the NL2SQL agent, we assume access to an LLM and an embedding model, which can be used to help with translation. We use the embedding model to embed strings and compute their cosine similarity when performing similarity search; search between strings $x$ and $x'$ is denoted by \texttt{emb-sim}($x$, $x'$). Additionally, the LLM is not an NL2SQL agent unlike $\mathcal{A}$ and does not have query access to the database.

\begin{table}[t]
\centering
\vspace{-2pt}
\footnotesize
\setlength{\tabcolsep}{4pt}
\renewcommand{\arraystretch}{1.1}
\caption{Notation.}
\label{tab:notation}

\begin{tabular}{l p{0.66\columnwidth}}
\toprule
\textbf{Symbol} & \textbf{Description} \\
\midrule
\multicolumn{2}{l}{\emph{Database and Query}} \\
\midrule
$\mathcal{D}$ & Relational database \\
$q$ & NL query \\
$s$ & SQL query; $s^\star$ denotes ground-truth SQL \\
\midrule
\multicolumn{2}{l}{\emph{NL2SQL Agent and Augmentation}} \\
\midrule
$t$ & Iteration index \\
$\mathcal{A}$ & NL2SQL agent \\
$C_t$ & Agent context at $t$ \\
$\tau$ & Agent execution trace \\
$\mathcal{E}$ & Experience set \\
$M$ & Augmentation mechanism \\
$C_t^M$ & Augmented context \\
$\mathcal{Q}$ & Dist. future NL queries \\
\bottomrule
\end{tabular}
\vspace{-6pt}
\end{table}

\begin{algorithm}[t]
\SetAlgoLined
\SetAlgoNoLine
\DontPrintSemicolon
\LinesNumbered
\caption{Agent Augmented with Mechanism $M$}
\label{alg:augmented_agent}

$C_0 \leftarrow q$\;
$t \leftarrow 0$\;
$C_0^M \leftarrow M(C_0,\mathcal{E})$\;
$\texttt{is\_final} \leftarrow \texttt{False}$\;

\BlankLine
\While{\texttt{is\_final} = \texttt{False}}{
  $(s_t,\ \texttt{is\_final}) \gets \mathcal{A}(C_t^M)$\;
  $R_t \leftarrow \textsc{ExecSQL}(s_t)$\;
  $C_{t+1} \leftarrow \texttt{concat}(C_t^M, R_t, s_t)$\;
  $C_{t+1}^M \leftarrow M(C_{t+1}, \mathcal{E})$\;
  $t$++\;
}
\BlankLine
\textbf{return} $s_{t-1}$\;
\end{algorithm}

\vspace{-4pt}
\subsection{Augmenting Agents using Experience}\label{sec:objective}
\textbf{Augmented Agents}. Our goal is to improve the accuracy of an NL2SQL agent by allowing the agent to accumulate experience when using the database. We consider the setting where the agent is initially given a few NL query examples with known ground-truth to accumulate experience, analogous to a ``training phase'' for an LLM. These examples are used to help answer future queries through an \textit{augmentation mechanism} that uses the experience to provide feedback to the agent. %To do so, our objective is to design a mechanism that steers the agent during SQL generation by injecting feedback into its context, based on experience accumulated from previous questions.
To formalize this process, first define an \textit{experience tuple} as a tuple $(q, \tau, s^\star)$, where $q$ is a natural-language query, $\tau$ is the agent’s execution trace when attempting to answer $q$, and $s^\star$ is a SQL query that produces the ground truth result. We assume access to a collection of $N$ past experience tuples,
\[
\mathcal{E} = \{ (q_i, \tau_i, s_i^\star); \forall i\in[N] \},
\] 
called an \textit{experience set}.
%where $q_i$ is a natural-language query, $\tau_i$ is the agent’s execution trace when attempting to answer $q_i$ (including intermediate SQL drafts and execution results), and $s_i^\star$ is a SQL query that produces the correct execution result for $q_i$ on the database.

To use the experience set when generating a new query, we define an \textit{augmentation mechanism} as one that that modifies the agent’s context during execution based on $\mathcal{E}$. Formally, let $C_t$ denote the agent’s context at step $t$. An augmentation mechanism is a function, $M$, which takes as input the past experience $\mathcal{E}$ and a context $C_t$, and produces a new context $C_t^M$ 
% \joey{The superscript $M$ could be a little confusing.  Maybe $\hat{C}_t$ or $C^\prime_t$? What do others think?}\joey{I changed my mind its probably fine.}
by optionally adding natural language feedback derived from $\mathcal{E}$ to $C_t$, i.e., 
\[
C_t^M=M(\mathcal{E}, C_t).
\]
Note that $C_t^M$ can simply contain experiences tuples appended to $C_t$, or---more beneficially as we show later---can contain knowledge distilled from the experience. This new context $C_t^M$ is then provided to the NL2SQL agent at step $t$ instead of $C_t$, producing a SQL query
\[
s_t, \texttt{is\_final} = \mathcal{A}(C_t^M).
\]
We say the mechanism $M$ \textit{augments} the agent $\mathcal{A}$ when it is used to modify its context from $C_t^M$ to $C_t$. Alg.~\ref{alg:augmented_agent} shows the process of using an augmented agent to perform translation. $M$ modifies the agent's context (Line 7) and $\mathcal{A}$ is called on this modified input context (Line 3). We call an agent that follows Alg.~\ref{alg:augmented_agent} an \textit{augmented agent}. 

\textbf{Problem Statement}. Our goal is to design an augmentation mechanism that uses past experience to improve the accuracy of the NL2SQL agent on future queries. Formally, let $\mathcal{Q}$ be the distribution of future NL queries.  
Then, given an agent $\mathcal{A}$ and an experience set $\mathcal{E}$, our goal is to design a mechanism $M$ that augments $\mathcal{A}$ (as in Alg.~\ref{alg:augmented_agent}) to improve its accuracy on queries from $\mathcal{Q}$. %the agent is invoked iteratively as described in Section~\ref{sec:definition}, with the steering mechanism $M$ applied at selected steps. 
Let $s(q, M)$ denote the translation result  $\mathcal{A}$ produces when augmented with a mechanism $M$. Our objective is to design an augmentation mechanism $M$ that maximizes execution accuracy on future queries:
\[
\max_M \;\; \mathbb{E}_{q\sim\mathcal{Q}}\big[
\mathbb{I}\!\left[
\textsc{ExecSQL}(s(q, M), \mathcal{D}) \equiv \textsc{ExecSQL}(s^*, \mathcal{D})
\right]\big],
\]
% \joey{Shouldn't the right hand side of the equivalence be $\textsc{ExecSQL}(s*, \mathcal{D})$ or $\textsc{ExecSQL}(s^*(q), \mathcal{D})$.  The gold truth shouldn't depend on the modification of the trace (or a trace at all)?}
where $\equiv$ denotes execution equivalence and $\mathbb{I}$ is indicator function.

\section{Augmenting NL2SQL Agents with Tribal Knowledge}
\label{sec:agents}

\begin{figure*}[t!]
    \centering
    \vspace{-3em}
    \includegraphics[scale=0.7]{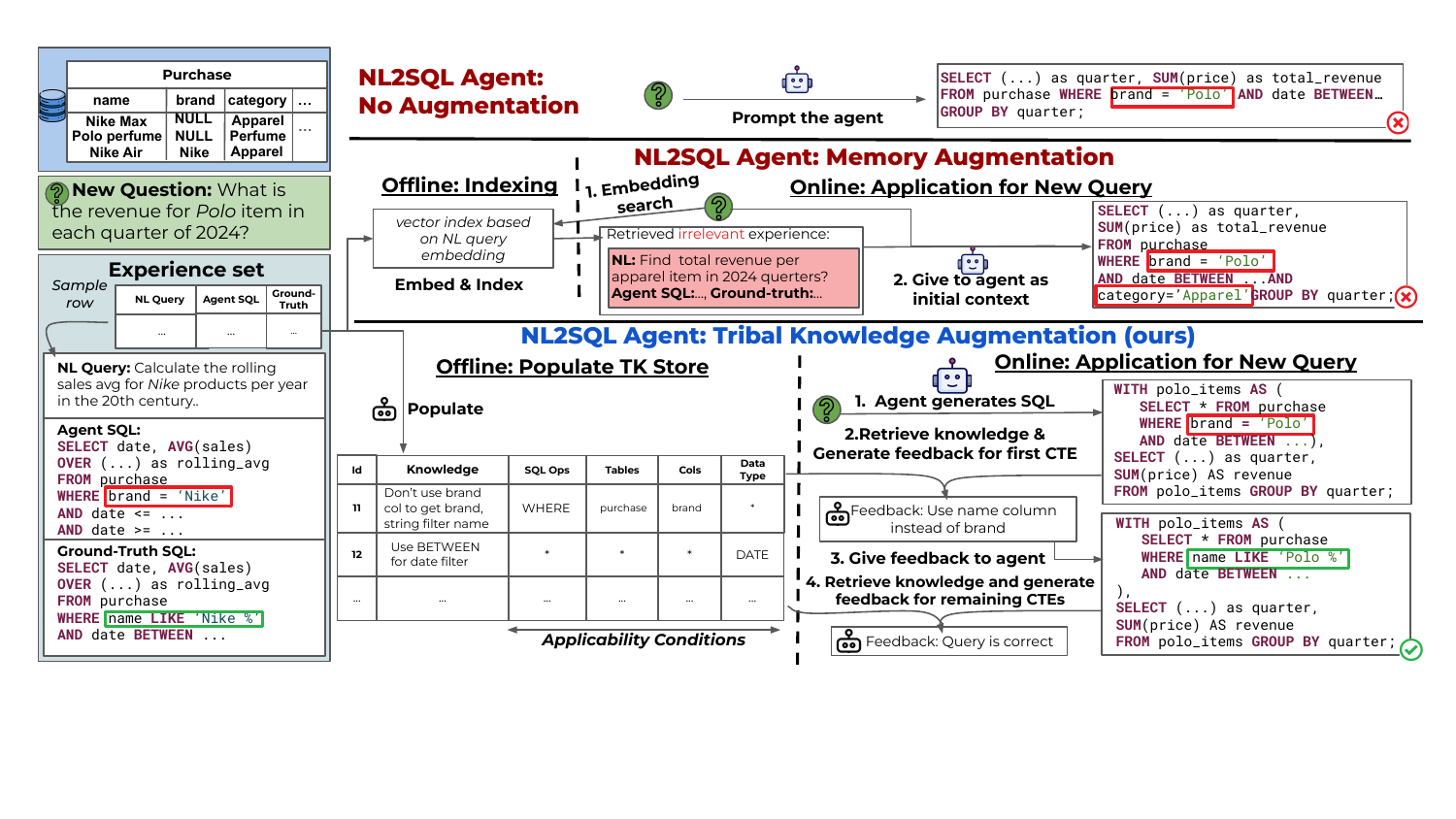}
    \vspace{-2cm}
    \caption{Augmenting NL2SQL agents with tribal knowledge. The un-augmented agent incorrectly uses the brand column to filter for Polo products. Adding memory augmentation retrieves irrelevant information, introducing a new error instead of fixing the existing one. \sys fixes the error using knowledge generated from past experience, filtering for Nike products.}
    \label{fig:overview}
\end{figure*}

We augment the NL2SQL agent with \emph{tribal knowledge} that enables the agent to accumulate and use knowledge derived from previous experiences to correctly query the database. We first introduce a simple memory augmentation solution in Sec.~\ref{sec:memory} to motivate our design. We then present an overview of our approach in Sec.~\ref{sec:tk_overview}, with details in Secs.~\ref{sec:tk_generation} and~\ref{sec:tk_application}.

%improve the accuracy of NL2SQL agents using  . Rather than replaying past examples verbatim, we design an augmentation mechanism that extracts reusable knowledge from an experience set and applies it as targeted feedback during future query generation. In this section, we first introduce a general framework for knowledge-augmented agents and describe a naive instantiation that highlights key challenges. We then provide an overview of our approach, \sys, which addresses these challenges through coordinated design choices across knowledge discovery, indexing, and application.

%\texttt{emb-sim} indicates the cosine similarity between the embeddings of the NL queries $q_i$ and $q$
\subsection{Naive Approach: Memory Augmentation}\label{sec:memory}
%\textbf{Naive Approach.}
A simple augmentation mechanism is to include relevant experience tuples in the agent's context, similar to how memory stores do so through a process of RAG~\cite{lewis2020retrieval}. Given a query $q$, we retrieve past experience tuples for similar queries as 
$$
\mathcal{E}_q = \operatorname*{top-k}_{(q_i,\tau_i,s_i^*) \in \mathcal{E}} \texttt{emb-sim}(q_i,q),
$$ 
where $\operatorname*{top-k}$ is the $k$ most similar experiences. We then use the augmentation mechanism $M$ to include $\mathcal{E}_q$ in agent's initial context: 
$$C_t^M=\concat(q, e_1, ..., e_k)\text{ when }t=0;\quad C_t^M=C_t\text{ otherwise},$$
where $e_i=\concat(q_i, \tau_i, s_i^*)$ corresponds to the $i$-th experience tuple $(q_i, \tau_i, s_i^*)$ in $\mathcal{E}_q$. %We call this approach \textit{preventative memory augmentation}, \textit{preventative} because it attempts to prevent the agent from making mistakes by inputting past experience tuples in the initial context and \textit{memory} because it simply allows the agent to remember past queries.  
To illustrate this approach, Fig.~\ref{fig:overview} revisits the example from Sec.~\ref{sec:intro}: querying a database that contains a \texttt{purchase} table where \texttt{brand} is inconsistently populated and instead \texttt{name} reliably includes product names and brands (top left). 
% \joey{I am trying to use our notation here, maybe also add it to the figure?}
The experience set $\mathcal{E}$ records a prior query where the agent incorrectly used \texttt{brand} instead of \texttt{name}.
% \joey{Instead of bottom left could we use something like \circledtext[height=1.9ex,charshrink=0.65]{1} or \ding{182} in the figure and text.}
When answering a new query (highlighted green), the agent without any augmentation mechanism repeats this mistake \cite{yao2023react}. Naive memory augmentation instead retrieves past experience tuples from a memory store (implemented as a vector index over query embeddings) and injects them into the agent's initial context, but this results in even more errors.

This approach has two main drawbacks. First, embedding similarity often retrieves irrelevant past experiences. In Fig.~\ref{fig:overview}, the un-augmented agent's error stemmed from misunderstandings about which column to use, and the retrieved experience---though highly similar to the new question---failed to address the misunderstanding because it did not involve the \texttt{name} or \texttt{brand} columns that caused the error. Second, the agent often misuses retrieved experience appended to its initial context, drawing incorrect conclusions that introduce new errors. In Fig.~\ref{fig:overview}, the agent wrongly adds a filter on the category column simply because it appeared in past experience.

\subsection{\sys: Tribal Knowledge Augmentation}\label{sec:tk_overview}
%To use this knowledge, rather than augmenting the model’s initial context, \sys uses a \textit{corrective augmentation} paradigm: it uses an augmentation mechaafter the agent generates a SQL query $s_t$, a mechanism $M$ checks whether it matches the intended semantics and, if not, augments the context $C_t$ with corrective feedback. This approach is more accurate because it detects and corrects errors in the generated SQL rather than predicting and preventing them in advance.

Past experience on its own is insufficient to correct the agent's logical errors when answering new queries. %; the agents cannot effectively produce correct  misconceptions by only observing the experience tuples. 
Instead, we propose to discover \textit{tribal knowledge} from the past experience and provide this knowledge to the agent when pertinent. % This knowledge needs to be provided to the agent when pertinent to the query,  % and avoid the pitfalls of the naive approach that 
To do so, we propose \sys; \sys discovers tribal knowledge describing how to correctly query the database and stores it in a \textit{Tribal Knowledge Store} (TK-Store) for retrieval. 
%discovers tribal knowledge about how to correctly query the database from the experience set and uses this knowledge to 
%and stores it in a \textit{Tribal Knowledge Store} (TK-Store) that ingests experience, discovers knowledge, and indexes it for retrieval. 
%We propose \sys, a framework that augments agents with tribal knowledge. 
To use this knowledge, instead of inputting information in the models' initial context to \textit{prevent} agent's mistakes as our naive approach does (Sec~\ref{sec:memory}), \sys uses \textit{corrective augmentation}, i.e., using tribal knowledge to \textit{correct} the agent's mistakes after they occur. That is, when the agent produces a SQL query $s_t$ at the $t$-th iteration, the augmentation mechanism, $M$, evaluates $s_t$ to find errors. If it finds errors, it augments the context $C_t$ to provide feedback for correcting the SQL. Corrective augmentation enables accurate retrieval and feedback because it takes the agent-generated SQL into account, compared with the naive approach that uses only the original NL query. %in a SQL that will be generated from the . %, but instead allows the mechanism $M$ to \textit{detect and correct} the errors by taking into account . 

Concretely, \sys consist of two phases: (1) populating TK-Store in an offline step and (2) using TK-Store to augment the agent for new queries. We describe the TK-Store interface and data model in Sec.~\ref{sec:tk_store_overivew}, overview how it's populated in Sec.~\ref{sec:overview:populate_tkstore}, and discuss its use to augment the agent in Sec.~\ref{sec:augmet_overview}.

%\sys uses this knowledge to identify and correct the agent's errors after it generates a SQL, rather than retrieving it upfront using embedding similarity of NL queries. This process allows for more accurate retrieval because \sys first observes the operations the agent performs and the data sources it accesses to answer the query, and retrieves knowledge that resolve the agent's misconceptions about using those operations and data sources. 

%\sys performs the above knowledge discovery, retrieval and application in two steps. First, in an offline step, \sys builds what we call a \textit{tribal knowledge store} that discovers knowledge from the experience set and indexes it for retrieval. 

%TK-Store is populated by ingesting agent's past experience 
\subsubsection{Tribal Knowledge Store. }\label{sec:tk_store_overivew} The Tribal knowledge Store (TK-Store) is a data structure that stores tribal knowledge and supports retrieval. Here, we describe its interface and data model. The implementation of its operations is discussed in Sec.~\ref{sec:tk_generation}.

\if 0
It is populated by (1) discovering tribal knowledge from our experience set, and (2) for each piece of knowledge, specifying \textit{applicability conditions}, conditions, used for knowledge retrieval, that specify whether a piece of knowledge can help address misconceptions in a given SQL.

To index this knowledge, \sys uses knowledge \textit{applicability conditions}, conditions that specify which queries a piece of knowledge is useful for based on the logical operations performed for the query, rather than relying on NL query embeddings that merely takes into account high-level query semantics. Indexing the knowledge based on logical operations performed allows the agent to retrieve 

Finally, instead of retrieving all the relevant knowledge upfront and providing it to the agent at once with the NL query, \sys uses knowledge to identify the agent's errors after it generates a SQL and provide fine-grained feedback to the agent to correct them. \sys does so by retrieving knowledge and providing feedback to the agent one sub-query at a time, ensuring fine-grained feedback that the NL2SQL agent can apply correctly. 
\fi

\textbf{Data Model}. TK-Store stores tribal knowledge; for each such tribal knowledge statement in NL, TK-Store stores \textit{applicability conditions}, which are features of a SQL query for which the knowledge is expected to help correct misconceptions. We consider four SQL features: \textit{SQL keywords} used in the query, \textit{table names} and \textit{column names} for tables and columns accessed, and \textit{data type} of the columns accessed in the query. Applicability conditions are used to retrieve knowledge statements. Informally, a knowledge statement is retrieved if a given SQL query's features \textit{match} the knowledge's applicability condition. A match occurs if the query contains the SQL keywords, tables, columns and/or data types included in the knowledge statement's applicability condition (formal discussion in Sec.~\ref{sec:knowledge-indexing}). %We select these features as they are the most common misconception sources for an agent.  %As such, we represent the applicability condition for a knowledge statement based on these query characteristics, and store the values for characteristics to enable retrieval. 
%Informally, a knowledge statement is retrieved if a given SQL query's characteristics \textit{match} the knowledge's applicability condition, e.g., if  where informally, a match occurs if the characteristics overlap with the applicability condition . %. These characteristics are used to retrieve relevant knowledge given a SQL query. 

Internally, knowledge statements and their applicability conditions are stored in a table. This table contains five columns: the knowledge statement, and the four features mentioned above. The knowledge statement is a string field, and the applicability conditions are stored as four columns each corresponding to a query feature, where each feature is stored as a set of strings. %Every tuple in this table contains two parts, a knowledge statement and an applicability conditions. 

Fig.~\ref{fig:overview} (bottom) shows TK-Store with two sample rows. The first row (id=11) contains knowledge about using  \texttt{name} column instead of \texttt{brand} when filtering products and the second about using BETWEEN keyword instead of a string filter for date values. Both knowledge statements have associated applicability conditions. For the first row, the applicability condition specifies that this knowledge is applicable to queries containing a WHERE clause on \texttt{brand} column in the \texttt{purchase} table. The value $*$ in a column implies that column can be ignored when evaluating whether the knowledge is applicable to a SQL query. For the second row the applicability condition states that the knowledge is applicable to queries that access a column with a date format.   %---we discuss the values shown and how they are used later in Sec.~\ref{sec:tk_generation}.

\textbf{Interface}. TK-Store supports two operations: \textsc{Insert} and \textsc{Retrieve}. $\textsc{Insert}(k, a)$ takes a knowledge statement, $k$, and its applicability condition, $a$, as input and inserts it into the TK-Store by simply creating a new row in the table. $\textsc{Retrieve}(s)$ is a function that takes as input a \textit{SQL query}, $s$, and outputs a subset of the knowledge statements in the TK-Store that can help resolve misconceptions in $s$. The \textsc{Retrieve} operation takes a SQL query as input to support corrective augmentation---that is, the \textsc{Retrieve} function is used after the agent has already generated a first attempt SQL query which is used to retrieve relevant knowledge. Details of the $\textsc{Retrieve}$ operations are presented in Sec.~\ref{sec:knowledge-indexing}.

%an applicability condition for a piece of knowledge $K$ is a tuple $A=$(\textit{SQL keywords}, \textit{table names}, \textit{column names}, \texttt{data type}) that specifies the knowledge $K$ is applicable to any SQL query if the SQL's characteristics \textit{match} those specified in the tuple $A$. We formalize the notion of a \textit{match} in Sec.~\ref{XXX}, but intuitively a SQL matches an applicability condition if it contains the SQL keywords, table names, column names or access columns with the specific \texttt{data type}.

%For example, for a knowledge $K$ containing information  that ``\texttt{name} column should be used instead of \texttt{brand} column to answer queries about product brand'', then a SQL query accessing the \texttt{brand} column may benefit from knowledge $K$. Thus, the applicability conditions for knowledge $K$ states that the knowledge is applicable for queries accessing \texttt{brand} column. 

%For each piece of knowledge, we additionally store knowledge \textit{applicability conditions}: conditions that specify SQL characteristics for which a piece of knowledge may be relevant. 

\vspace{-1em}

\subsubsection{Populating TK Store}\label{sec:overview:populate_tkstore} We define a function $\textsc{Populate}(\mathcal{E})$ that takes an experience set $\mathcal{E}$ as input, discovers tribal knowledge and applicability conditions based on the experience set, and inserts them into the TK-Store. It does so by observing the logical errors in the experience set and generating reusable knowledge statements to help correct them, detailed in Sec.~\ref{sec:knowledgegen}. Fig.~\ref{fig:overview} shows an example where multiple rows are generated from an experience tuple shown in the figure. The agent-generated SQL in the experience tuple uses an incorrect column (\texttt{brand} instead of \texttt{name}) and the population process generates a knowledge statement specifying that \texttt{name} should be used to correct the logical error.

\if 0
In this step, we first use the experience set to discover knowledge that can help address the agent's misconceptions. This is done by analyzing the experience set as well as the database to generate natural language statements that can help address the agents misconceptions. Then, to enable accurate knowledge retrieval, we introduce knowledge \textit{applicability conditions}, conditions that specify whether a piece of knowledge can help address misconceptions in a generated SQL.  Our tribal knowledge store is a data structure that (1) stores the knowledge and it's applicability conditions and (2) supports retrieving relevant knowledge for a SQL query (or sub-query) using applicability conditions.
\fi

%\textbf{Populating TK-Store and Retrieval}. The operation \texttt{Populate} populates the TK-Store, using the experience set to discover tribal knowledge. \sys does so by observing the mistakes the NL2SQL agent makes in the experience set, exploring the data and creating NL statements that address the NL2SQL agent's misconceptions. Then for for each generated knowledge, \sys generates the applicability condition based on the knowledge itself as well as the experience tuple the knowledge was derived from. TK-Store implements the \texttt{Retrieve} function by evaluating whether a given SQL query matches the applicability conditions for each knowledge and returns a subset of the knowledge that are deemed relevant based on the applicability conditions. Details of population and retrieval are presented in Sec.~\ref{sec:augmet_overview}. 

\subsubsection{Augmentation Mechanism Overview. } \label{sec:augmet_overview}
Using \textit{corrective augmentation}, given a new query, \sys first asks the NL2SQL agent to generate a first attempt SQL to answer the query. \sys retrieves knowledge based on this first attempt. The retrieved knowledge is converted into NL feedback and provided to the NL2SQL agent to fix any potential mistakes in the query. Since SQL queries in practice typically
% are complex and 
contain multiple sub-queries, the NL2SQL agent often fails to correct errors if given feedback for the entire query at once. Instead, \sys provides feedback one sub-query at a time, allowing for more fine-grained steering of the NL2SQL agent (as described below). 

%\joey{I broke the previous paragraph into two and more clearly restated the transition describing the why of the contents.  We may want to apply this process throughout the paper.}
% To provide feedback at each step,  \sys instructs the agent to use Common Table Expressions (CTEs) to write queries. It then uses the tribal knowledge to identify misconceptions, one CTE at a time. For each CTE, it retrieves relevant knowledge from the TK-Store, and based on the retrieved knowledge, the database, and the original CTE, \sys provides feedback to the NL2SQL agent. After the agent fixes the CTE, \sys moves to the next CTE, providing feedback to the NL2SQL agent until all the CTEs are corrected.  

To do so, \sys instructs the agent to use Common Table Expressions (CTEs) to write queries. It uses tribal knowledge to identify misconceptions, one CTE at a time. Per CTE, it retrieves relevant knowledge from the TK-Store, and based on the retrieved knowledge, the database, and the original CTE, \sys issues feedback to the NL2SQL agent. After the agent fixes the CTE, \sys moves to the next CTE, providing feedback to the NL2SQL agent until all the CTEs are corrected.

Fig.~\ref{fig:overview} (bottom right) shows an example of this process. First, the agent produces a SQL containing a CTE that incorrectly uses the column \texttt{brand}. \sys uses the CTE to retrieve knowledge from the TK-Store and generate feedback that the column \texttt{name} should be used instead of \texttt{brand} (Step 2). %Note that \sys retrieves the knowledge (with id=11) that \texttt{name} column should be used instead of \texttt{brand} because this knowledge has applicability condition that specifies it is relevant for all queries that filter based on \texttt{brand} column. 
\sys then provides this feedback to the NL2SQL agent (Step 3), which then corrects the CTE. \sys finally studies the other sub-queries generated, retrieves knowledge and observes that the query is correct, providing this feedback to the agent which returns the final correct query.

\section{TK-Store Population and Retrieval}\label{sec:tk_generation}
We now describe how \sys populates TK-Store from past experience sets and how TK-Store perform retrieval. We discuss TK-Store population in Sec.~\ref{sec:knowledgegen} and retrieval in Sec.~\ref{sec:knowledge-indexing}.

\subsection{TK Store Population}\label{sec:knowledgegen}

% \joey{break up this paragraph and ensure that first sentence is the key idea of the paragraph. Ideally start this section with restating the challenge. -- I split off the first paragraph but the second may need some work.}
The $\textsc{Populate}(\mathcal{E})$ function is responsible for extracting tribal knowledge and applicability conditions from the experience set $\mathcal{E}$ and populating the TK Store.
% takes an experience set $\mathcal{E}$ as input and populates the 
% TK-Store by discovering knowledge and applicability conditions from $\mathcal{E}$ and inserting them as new rows in the TK Table. 
A simple method to do so is by directly asking an LLM to analyze each experience tuple, $(q, \tau, s^*)\in\mathcal{E}$ and identify important facts about the database or how to answer future queries correctly.
However, directly extracting facts from each experience can result in (1) incorrect or trivial knowledge and (2) knowledge specific to $q$ that cannot be reused.
This is because the LLM is often unable to detect logical errors in the agent-generated SQL, especially when it is complex and errors depend on subtle data characteristics.
% , and (2) knowledge specific to $q$ that cannot be reused for future queries.  
%to ensure accurately identifying the logical errors and generating statements that can correct them, 

Instead, \sys first identifies the logical errors and generates \textit{correction statements}---NL statements that correct the logical errors through an iterative process that explores the database and analyzes differences between the agent SQL and ground-truth SQL until all logical errors are correctly identified. 
%Instead, \sys first accurately identifies the logical errors and generates \textit{correction statements}, specific NL statements that can correct the logical errors, through an iterative process that explores the database and studies various differences between the agent-generated and the ground-truth SQL until it hones in on the logical errors. %  various makes small modifications to the query to enable more accurate identification of the logical error.  
\sys then converts these correction statements into reusable knowledge.
Each correction statement addresses a logical error specifically for the SQL query it was designed to fix; \sys identifies other queries where the same logical errors can occur and generates a knowledge statement to correct them, ensuring reusability. 
The applicability conditions are generated similarly, by identifying features of SQL queries where the logical errors may appear. 
This two-step process is shown in Alg.~\ref{alg:populate} (throughout this paper, functions in \textcolor{blue}{blue} are LLM pipelines, and prompts are available in the supplementary material). \sys (1) generates correction statements that help correct logical errors in each experience tuple  (Line~\ref{line:populate:get_error}) and then (2) based on each correction statement, generates generalizable knowledge statements and their applicability conditions  (Line~\ref{line:populate:getrow}) which are then inserted as a row in TK-Store.
\subsubsection{Generating Correction Statements. }\label{sec:find_logical_diff}
\begin{algorithm}[t]

\small
\DontPrintSemicolon
\SetAlgoLined
\LinesNumbered

\caption{\textsc{Populate}$(\mathcal{E})$}
\label{alg:populate}

%\textbf{Input:} NL question $q$; database $\mathcal{D}$; initial SQL $s_0$; ground truth SQL $s^\star$; tribal knowledge store $TK$\\
\vspace{5pt}

%\SetKwProg{Fn}{Procedure}{:}{}
%\Fn{\textsc{\textup{Populate}}$(\mathcal{E})$}{
\label{alg:gen:beginpop}
    \ForEach{$e \in \mathcal{E}$}{
        $\mathcal{K}\leftarrow \textsc{GetCorrectionStatements}(e)$\label{line:populate:get_error}
        
        \ForEach{$k \in \mathcal{K}$}{
            $(k, a) \gets \textcolor{blue}{\textsc{GenTKRow}}(q,\mathcal{D},k)$\label{line:populate:getrow}
            
            TK-Store.insert($k, a$)
            %$\widetilde{\mathcal{K}} \gets \widetilde{\mathcal{K}} \cup \{(k,a)\}$\;    
        }   
    }
    \label{alg:gen:endpop}
%}
\end{algorithm}

\begin{algorithm}[t]
\small
\DontPrintSemicolon
\SetAlgoLined
\LinesNumbered

\caption{\textsc{GetCorrectionStatements}$(e)$}
\label{alg:knowledge-generation}
\SetKwInOut{Input}{Input}
\SetKwInOut{Output}{Output}

\Input{Experience tuple $e$}%; database $\mathcal{D}$; initial SQL $s_0$; ground truth SQL $s^\star$; tribal knowledge store $TK$\\
\Output{Correction statements for logical errors in $e$}

    $(q, \tau, s^*)\leftarrow e$, $s\leftarrow $ incorrect SQL in $\tau$
    
    $\Delta \gets \varnothing$
    
    $R^\star \gets \textsc{ExecSQL}(\mathcal{D}, s^\star)$,\,\quad $R \gets \textsc{ExecSQL}(\mathcal{D}, s)$\;
    
    \While{$R \not\equiv R^\star$}{
        $(\delta, s) \gets \textcolor{blue}{\textsc{MakeCorrection}}(q,\mathcal{D}, s, s^\star, R, R^\star)$\label{alg:correction:make_correction}\;

        \if 0
        $\texttt{agent\_sim} \gets \textsc{Jaccard}(s', s)$\;
        $\texttt{gt\_sim} \gets \textsc{Jaccard}(s', s^\star)$\;
    
        \If{
            $\texttt{agent\_sim} \ge 0.3$\
            $\wedge$\ $\texttt{gt\_sim} > 0.7$\
            $\wedge$\ \\ $\texttt{gt\_sim} > \texttt{agent\_sim} + 0.2$
        }{
            \textbf{continue} \tcp*{reject near-ground-truth rewrite}
        }
        \If{
            $\textsc{Jaccard}(s^*, s')-\textsc{Jaccard}(s, s')>\alpha$\label{alg:correction:if_sim}
        }{
            \textbf{continue} \tcp*{reject near-ground-truth rewrite}\label{alg:correction:if_sim_end}
        }
        \fi

        $R \gets \textsc{ExecSQL}(\mathcal{D}, s)$
    
        $\Delta \gets \Delta \cup \{(\delta, s, R)\}$\;\label{alg:correction:delta}
    }
    \textbf{return} $\textcolor{blue}{\textsc{GetRequiredCorrections}}(\Delta)$\label{alg:correction:condense}
%\tcp*[r]{select logical diffs}
\end{algorithm}

% \joey{Maybe rewrite: To generate correction statements we need to ... then you can write "In this step....}
To generate correction statements, we need to first find logical errors in the agent's SQL. Concretely, given an experience tuple, $e=(q, \tau, s^*)$, where $s$ is the agent-generated SQL query ($s$ is included in $\tau$), we find the logical errors in $s$ and generate correction statements that can correct these errors by comparing $s$ with $s^*$. We generate separate correction statements for each logical error in $s$, generating \textit{atomic} correction statements %(and thereby knowledge generated from them, as we see in Sec.~\ref{sec:genknowledge}) 
that can be combined together to correct different combinations of logical errors---we say a correction statement is \textit{atomic} if it cannot be decomposed into multiple semantically meaningful statements that each address logical errors.   

\textbf{Algorithm}. To accurately identify logical errors and generate correction statements, \sys first identifies a set of \textit{candidate corrections} that are atomic and, together, address the logical errors in the SQL; from these candidate corrections, \sys removes incorrect or redundant corrections to obtain the final set of correction statements. While creating the candidate corrections set, we iteratively ask an LLM to explore the database and modify a single SQL clause (e.g., \texttt{WHERE} clause, \texttt{GROUP BY} expression) %, or a single CTE body) 
in the agent-generated SQL, by both producing a new SQL query and an NL statement that discusses the correction applied. We only allow single clause modifications to ensure each correction is atomic. This process continues until we obtain a SQL query that produces the correct answer---we use equivalency with the ground-truth result set as a proxy for correctness---ensuring the logical errors in the SQL are fixed through these sequence of corrections. Nonetheless, not every correction made by the LLM may be correct or necessary. Thus, after all the corrections are performed, we make another LLM call to review its previous work and identify a subset of the candidate corrections that are needed to correct the logical errors following a self-reflection approach \cite{shinn2023reflexion}. 

% \joey{Can we claim to be using "self-reflection" or any other standard agent design patterns?}

Alg~\ref{alg:knowledge-generation} details how we obtain the set of correction statements. Given an experience tuple, we iteratively ask an LLM to perform a correction for a single clause in the SQL (Line~\ref{alg:correction:make_correction}).
The LLM outputs a new SQL query and an NL description of the correction applied, $\delta$. All the corrections (the NL statements along with the corresponding SQL draft) are kept in the set $\Delta$ (Line~\ref{alg:correction:delta}). Finally, we ask another LLM to evaluate the full set of corrections and keep only those required to fix the logical errors (Line~\ref{alg:correction:condense}).

\begin{figure}[t]
    \centering
    \vspace{-1em}
    % solid black square, 1cm x 1cm
    \includegraphics[scale=0.45]{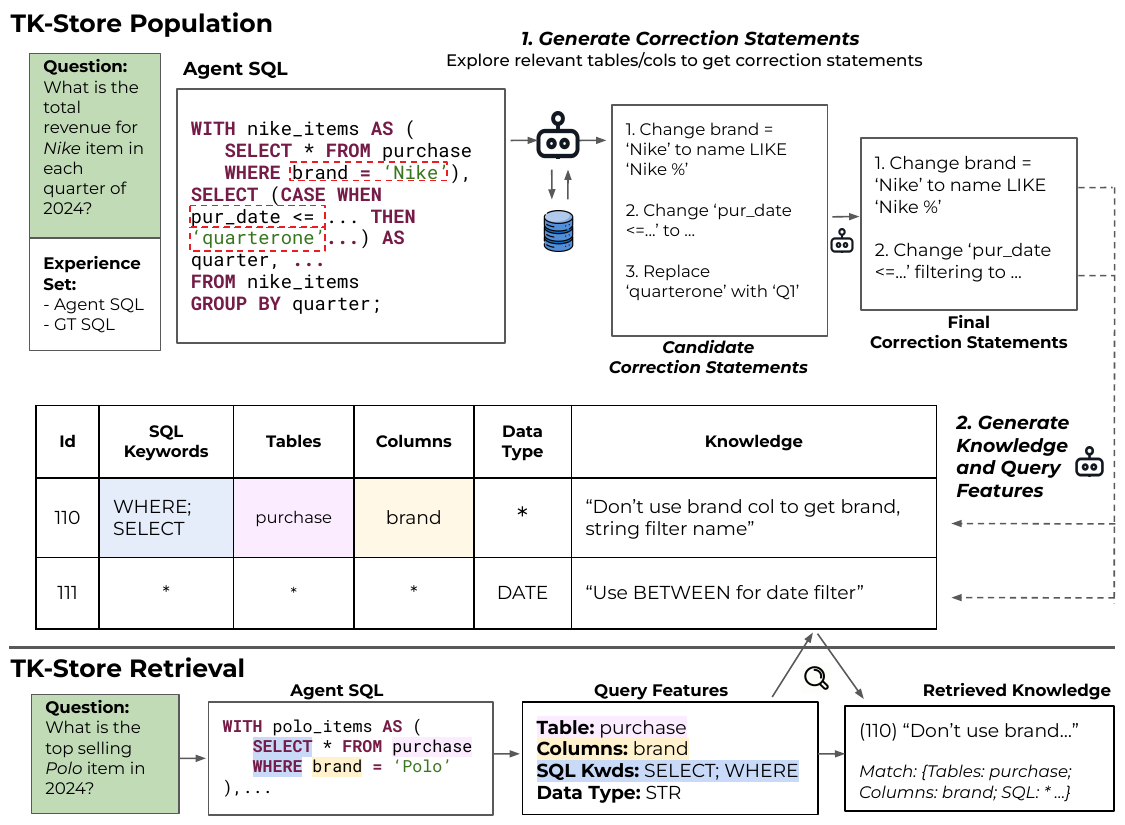}
    \caption{Knowledge store population (top) and retrieval (bottom). Features from input SQL are matched against knowledge applicability conditions in the TK-Store for retrieval.
    }
    \label{fig:knowledgestore}
\end{figure}

\textbf{Example}. Fig.~\ref{fig:knowledgestore} (top) illustrates this process on our example question. Given the agent-generated SQL, \sys first generates a list of candidate correction statements to fix the query, done iteratively while exploring relevant tables and columns and taking the ground-truth SQL into account. Here, the LLM generates three candidate correction statements, one of which is to replace the \texttt{brand} filter with a string filter on \texttt{name} column---crucially, data exploration helps the LLM realize the \texttt{brand} column contains missing values while the \texttt{name} column reliably contains product brands. From the three candidate correction statements, the LLM removes the renaming of the column ``quarterone'' to ``Q1'' (that did not contribute to a logical error) to create a final set of correction statements. %Finally, in the second step in Fig.~\ref{fig:knowledgestore} (middle), an LLM call is used on the final correction statements to create generalized knowledge. The date filtering correction statement in the example does not apply to only the \texttt{'pur\_date'} column, so the LLM generated knowledge statement abstracts the column and table names before inserting the knowledge statement in the store.

\vspace{-0.5em}
\subsubsection{Generating Knowledge and Applicability Condition}\label{sec:genknowledge} 
For each correction statement created in the previous step, we generate knowledge statements and applicability conditions jointly through a single LLM call to ensure they remain consistent. 
%Generating Knowledge and applicability conditions is done through an LLM call, where we instruct an LLM to generate knowledge statements that correct the given logical error and applicability conditions for the knowledge. We generate the
%a lesson to be taken from each edit in the minimal edit set. To maximize \textit{reusability}, we instruct that data-usage knowledge statements should obscure specific table and column names. 
% \joey{Don't start with recall.  First sentence should say what is a knowledge statement and what it does.  Essentially all paragraphs should start with a single statement that embodies the key point of that paragraph.  AI can probably help you here.}

\textbf{Knowledge Statement}. We instruct the LLM to use each correction statement, $\kappa$, to generate a knowledge statement that fixes the logical error, $l$, that the correction statement originally addressed. The knowledge generated is reusable, as it fixes $l$ in all future queries where it may occur. In practice, this reusability is achieved by removing query specific details when possible, i.e., removing dependencies on column and table names, data types, or column values, so that the knowledge is applicable more broadly. Fig.~\ref{fig:knowledgestore} shows an example of this process where knowledge statements are created from correction statements (in Step 2), showing how the knowledge statements generalize correction statements. For Statement 1 in the final correction statements set, we see that the correction statement only applies to queries asking for Nike products, but knowledge generation creates a knowledge statement that applies to any query looking for product brands.  Fig.~\ref{fig:knowledgestore} shows another example where the correction statement (Statement 2) specifies the column \texttt{pur\_date}, but knowledge generation creates a more general statement that applies to any column with date information.  

%\sep{modify this and point out how we generalize to more sub-tasks}Figure~\ref{fig:knowledgestore} illustrates knowledge generation examples. . Observe that the previous step detected two logical errors in the agent-generated SQL. In this step, we generate two knowledge statements with their applicability conditions to correct the logical errors. The first is designed to correct the logical error that name column should be used instead of brand. The second corrects the error that date filtering should be done using BETWEEN.   %The LLM is instructed to generate knowledge statements that can correct the logical error  to as many sub-tasks as possible based on the takeaways from the correction statement. , is created by removing de
% lead to  Recall that an applicability condition for a knowledge statement specifies characteristics of a SQL query  for which the knowledge is beneficial. 
% \joey{this previously started with a recall.  I changed to a statement of what you should know but this paragraph should really start with a statement about Applicability Conditions.  As in Applicability conditions determine ...}

\textbf{Applicability Conditions}.  An applicability condition for a knowledge statement specifies possible features of an incorrectly generated SQL where the agent has misconceptions. Using the applicability condition, retrieving the knowledge statement for future SQL queries with the same features becomes more reliable than semantic similarity search.

% To obtain applicability conditions for a knowledge statement, the LLM is instructed to predict the ways a SQL query may incorrectly perform the query that the knowledge statement is designed for. The LLM is then asked to determine the features of the logically incorrect clauses in those SQL queries; specifically, what SQL keywords they may contain, what tables and columns they may access and for what data types the errors may occur. Specifically let $\mathcal{X}= \{\texttt{SQL\_keywords}, \texttt{tables}, \texttt{columns}, \texttt{data\_types}\}$, be the set of four characteristics we consider. Then, for each $x\in \mathcal{X}$, the LLM outputs either a list of strings or a wildcard ($*$). Unless the output is a wildcard, for \texttt{SQL\_keywords}, the LLM is only allowed to output SQL keywords (e.g., SELECT, WHERE, JOIN, GROUP BY, etc. ), for \texttt{tables} and \texttt{columns} allowed to output existing table and column names in $\mathcal{D}$ and for \texttt{data\_type} only a valid SQL data type (e.g., int, DATE). For any feature in $\mathcal{X}$, the LLM is instructed to specify a value only if it must exist in an incorrect SQL; otherwise it  assigns a wildcard ($*$). 

To obtain applicability conditions, the LLM is instructed to predict how a SQL query may be incorrectly formed for the target task, and to identify the features of the erroneous clauses—specifically, the SQL keywords involved, the tables and columns accessed, and the data types on which errors may occur. 

Let $\mathcal{X}=\{\texttt{SQL\_keywords}, \texttt{tables}, \texttt{columns}, \texttt{data\_types}\}$ denote these four features. For each $x \in \mathcal{X}$, the LLM outputs either a list of values or a wildcard ($*$). Unless a wildcard is used, \texttt{SQL\_keywords} must be valid SQL clauses (e.g., SELECT, WHERE, JOIN), \texttt{tables} and \texttt{columns} must exist in $\mathcal{D}$, and \texttt{data\_types} must be valid SQL types (e.g., \texttt{int}, \texttt{date}). A feature is specified only if it must appear in an incorrect SQL; otherwise, a wildcard ($*$) is assigned.

%The LLM output is a set of values for each query feature. % based on whether a corresponding SQL component is required for knowledge statement to apply.
%More generally, an applicability field is constrained if and only if the corresponding SQL component is required for the correction to apply. This ensures that applicability conditions generalize safely across unrelated query contexts. A query characteristic may take a wildcard value ($*$), indicating that the knowledge does not depend on that aspect of the query and should match broadly across contexts.

Applicability conditions can specify when a knowledge statement applies, either to operations on particular tables or columns, or to operations on a specific data type. The first row in Fig.~\ref{fig:knowledgestore} illustrates the former, where the knowledge about using \texttt{name} instead of \texttt{brand} applies to queries that refer to the \texttt{brand} column. The second row illustrates the latter, where the knowledge applies to any operation involving a date column.

\begin{algorithm}[t]
\small

\DontPrintSemicolon
\SetAlgoLined
\LinesNumbered
\caption{\textsc{Retrieve}$(s)$}
\label{alg:retrieve}

\textbf{Input:} SQL query $s$ \\
\textbf{Output:} Applicable knowledge statements $\mathcal{K}_{\text{ret}}$ \\
\vspace{6pt}
$c \gets$ parse $s$ and obtain query features\;\label{agl:retrive:get_char}

\vspace{4pt}
$\mathcal{K}_{\text{cand}} \gets \varnothing$\;
\label{alg:retrive:get_matches_begin}
\ForEach{\textnormal{row} $r$ \textnormal{\textbf{in}} \textnormal{TK-Store}}{
    \If{
        $\bigwedge\limits_{\forall x \in \mathcal{X}}\big((r[x] = *) \;\lor\; (r[x] \cap c[x] \neq \varnothing)\big)$\label{alg:retrieve:ismatch}
    }{
        $\mathcal{K}_{\text{cand}} \gets \mathcal{K}_{\text{cand}} \cup \{r[\texttt{knowledge}]\}$\;
    }
\label{alg:retrive:get_matches_end}
}

\vspace{4pt}
\tcc*[h]{\textbf{Context-aware filtering}}
\label{agl:retrive:ranking}
$\mathcal{K}_{\text{ret}} \gets \textcolor{blue}{\textsc{FilterKnowledge}}(s, \mathcal{K}_{\text{cand}})$\;

\Return $\mathcal{K}_{\text{ret}}$\;

\end{algorithm}

\if 0
$\textsc{GetQueryChars}(q,s,\mathcal{D})$ extracts query characteristics
$x = \{x[c]\}_{c \in \mathcal{C}}$ using an LLM; \\
$\textsc{RankKnowledge}(q,s,\mathcal{K})$ scores and orders knowledge statements by relevance using an LLM.
\fi

% \vspace{-1.5em}
\subsection{TK Store Retrieval}
\label{sec:knowledge-indexing}
At test-time, given a new query $q$ and a SQL draft $s$, $\textsc{Retrieve}(s)$ returns the relevant knowledge statements that can help correct errors in $s$ by first finding those whose applicability conditions \textit{match} $s$, and then using an LLM to filter the matching knowledge statements and keeping only the statements needed to correct the mistakes. 

%\subsubsection{Retrieving Knowledge}
%\label{sec:retrieve} We extract query characteristics for $s$ using the same procedure as used during knowledge generation.
% \alvin{why (only) these 4? e.g.. what about literals} re: tried to answer this above in Sec 3 Data Model. basically these are the features we choose to build our store since they are the most common sources of error that persist across queries (e.g. a future query is less likely to interact with the exact same literal than a table or column)

\textbf{Algorithm}. Alg.~\ref{alg:retrieve} describes the process for retrieving knowledge statements given a SQL query $s$. We first extract query features from $s$ using a pipeline with a SQL parser (Line~\ref{agl:retrive:get_char}), which extracts SQL keywords, tables, and columns that appear in the query $s$, and for each column we identify its data type by querying the database. The parser returns a dictionary $c$, such that $c[x]$ contains a set of values for each specific query feature $x\in\mathcal{X}$. Then, the algorithm finds knowledge statements by finding rows from the TK-Store whose applicability condition \textit{matches} those of the query features (Lines~\ref{alg:retrive:get_matches_begin}-\ref{alg:retrive:get_matches_end}). An applicability condition matches a query feature $c$ if for each query feature, the applicability condition contains that feature or is a wildcard (Line~\ref{alg:retrieve:ismatch}), that is, if
$$\bigwedge\limits_{\forall x \in \mathcal{X}}\big((r[x] = *) \;\lor\; (r[x] \cap c[x] \neq \varnothing)\big)$$
Finally, all knowledge statements whose applicability condition matches the SQL query are filtered by an LLM alongside the full SQL query to remove the knowledge statements deemed irrelevant.

%For each query characteristic value comprising the applicability condition for a knowledge entry, $s$'s corresponding query characteristic is matched against it. \textcolor{red}{match if either the stored value contains the query metadata or is a wildcard.} Only knowledge entries for which applicability condition is satisfied are retrieved. The applicability condition is satisfied if every query characteristic value $c_i$ is either a wildcard or intersects with the query characteristics in the $s$. 

% A metadata field is considered a match if either the stored value contains the query metadata or is a wildcard.  
% Only rules whose metadata matches across all indexed fields are retrieved.

% $\mathcal{I}$ over the knowledge store $\mathcal{K}$.
% \[
% \mathcal{I}(q, s)
% =
% \{\, r \mid (r, m_r) \in \mathcal{K} \;\wedge\; m_r \text{ matches } m(q, s) \,\}.
% \]

% \textbf{Example}.Fig.~\ref{fig:knowledgestore} (bottom) shows an examples of the retrieval process. Where given a SQL query as input, the retrieval process first identifies the query features and return knowledge statements whose applicability conditions match the query features. We see the query features match the SQL keywords, tables and columns in the first knowledge statement (id=110) so this knowledge statement is retrieved. 
\textbf{Example}. Fig.~\ref{fig:knowledgestore} (bottom) illustrates how, given an SQL query, retrieval first identifies its query features and returns knowledge statements whose applicability conditions match those features. Here, the query features match the SQL keywords, tables, and columns of the first knowledge statement (id=110), and therefore the statement is retrieved.

\begin{figure}[t]
    \centering
    \vspace{-1em}
    % solid black square, 1cm x 1cm
    \includegraphics[scale=0.475]{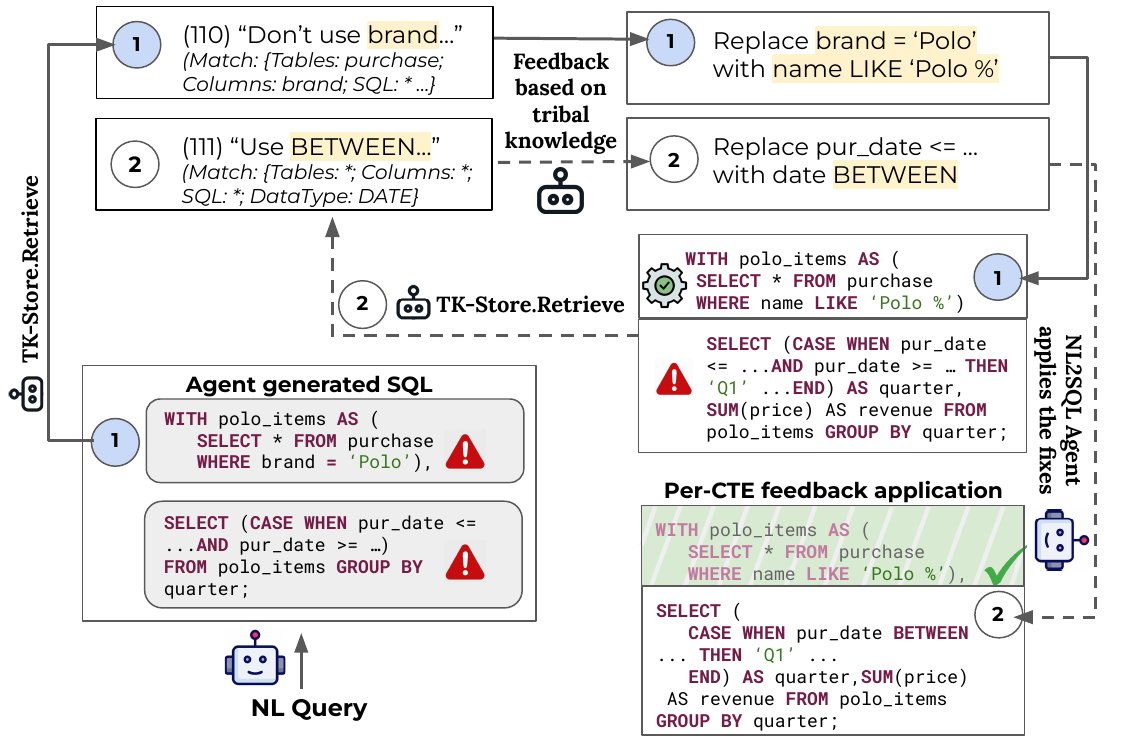}
    \caption{CTE-level application of tribal knowledge. LLM converts retrieved knowledge into localized feedback and applies corrections to CTE with which knowledge was retrieved, enabling precise refinement.}
    \label{fig:cte}
\end{figure}

\vspace{-0.5em}
\section{Augmenting Agents using TK Store}\label{sec:tk_application}
Our augmentation mechanism specifies \emph{when} to modify the agent’s context as well as \emph{how} to modify it.   
% \joey{don't tell me what you are going to do just do it.  The augmentation mechanism is resposible for ...}
As discussed in Sec.~\ref{sec:tk_overview}, instead of using the experience set to modify the agent's initial context, \sys uses \textit{corrective augmentation} that first asks the agent to generate a SQL query and uses the TK-Store to provide feedback to the agent to correct any errors.

\textbf{Augmentation Mechanism}. We modify the agent's context after it generates its first query attempt and provide feedback to ensure correctness (Alg.~\ref{alg:knowledge_application}). To detect whether the agent has produced a complete SQL query (the agent can issue intermediary SQL queries, as discussed in Sec.~\ref{sec:prelims}), the agent outputs a Boolean indicator, \texttt{is\_final}, indicating query completion. The agent’s context is modified only after \texttt{is\_final}=\texttt{True}, as shown in Alg.~\ref{alg:knowledge_application} Lines~\ref{alg:augment:is_final}–\ref{alg:augment:get_feedback}.

%Algorithm~\ref{alg:knowledge_application} describes procedure used to apply retrieved knowledge. As 
%discussed in Sec~\ref{sec:retrieve}, knowledge is retrieved using a test-time query $q$ along with a SQL draft with which we can search for relevant knowledge indexed on SQL contexts (e.g. SQL operators, tables, etc). As a result, we must apply our knowledge after an initial SQL draft is produced by the NL2SQL agent. 
%We modify Algorithm~\ref{alg:augmented_agent} by asking the NL2SQL agent to output an \texttt{is\_final} boolean after each step, signaling whether or not it is finished translating the query $q$. After the \texttt{is\_final} signal is produced, we apply our \emph{augmentation mechanism} to correct the initial SQL attempt.
% \joey{Does this sentence summarize this paragraph?}

% The way the augmentation mechanism modifies the context is shown in 

\begin{algorithm}[t!]
\small
\DontPrintSemicolon
\SetAlgoLined
\LinesNumbered

\textbf{Input:} NL query $q$; database $\mathcal{D}$\\
\textbf{Output:} Final SQL query $\hat{s}$.

\vspace{4pt}
\tcc*[h]{\textbf{NL2SQL Agent Loop}}

$C_0^{TK} \gets q$;\quad $t \gets 0$; $i \gets 0$\;

$\texttt{is\_final} = \texttt{False}$

\While{$\texttt{is\_final} = \texttt{False} \;\lor\; f \neq \varnothing$}{
    $f \gets \varnothing$\;
    $(s_t,\ \texttt{is\_final}) \gets \mathcal{A}(C_t^{TK})$\;
    
    \If{$\texttt{is\_final} = \texttt{True} \;\wedge\; i < \texttt{number\_of\_CTEs\_in}(s_t)$}{\label{alg:augment:is_final}
        \tcc*[h]{\textbf{Get Feedback on $i$-th CTE Using TK Store}}
        
        $f \gets \textsc{GetTKStoreFeedback}(s_t,\ i,\  \mathcal{D})$\;
        $i \gets i + 1$\;
        \label{alg:augment:get_feedback}
    }

    $R_t \gets \textsc{ExecSQL}(s_t,\ \mathcal{D})$\;
    $C_{t+1}^{TK} \gets \texttt{concat}(C_{t}^{TK},\ s_t,\ R_t,\ f)$\label{alg:augment:context}\;
    $t \gets t+1$\;
}
\Return $s_t$\;

\vspace{6pt}
\tcc*[h]{\textbf{Generate Feedback Using TK Store}}

\SetKwProg{Fn}{Procedure}{:}{}
\Fn{\textnormal{\textsc{GetTKStoreFeedback}}$(s,\ i,\ \mathcal{D})$}{
\label{alg:aug:M_begin}
    $(c_1,\dots,c_L)\leftarrow$ List of all CTEs in $s$\;

    $\mathcal{K} \gets \text{TK-Store}.\textsc{Retrieve}(c_i)$\;\label{alg:aug:retrieve}

    \Return $\textsc{\textcolor{blue}{Feedback}}(q,\ c_i,\ \mathcal{K})$\label{alg:aug:feedback}
    \label{alg:aug:M_end}
}

\caption{\textsc{Agent Augmented with TK Store}}
\label{alg:knowledge_application}
\end{algorithm}

% Given the SQL draft $s$, \sys retrieves knowledge for $s$ and turns it into feedback $f$ to append to the agent's context (Lines~\ref{alg:aug:M_begin}-\ref{alg:aug:M_end}). 
% A key observation is that different sub-queries within a single SQL query may contain different misconceptions, and correcting them therefore requires different knowledge statements. To improve precision in retrieval and knowledge application, we retrieve knowledge and provide feedback to the agent one CTE at a time (we instruct the agent in its initial context to break down its SQL to as many CTEs as possible) For simple translations or instances where the agent writes SQL with no CTEs, we treat the entire SQL as one CTE for our method. For each CTE, we first retrieve relevant knowledge (Line~\ref{alg:aug:retrieve}) and then use an LLM to summarize all the knowledge into actionable feedback to correct the specific CTE (Line~\ref{alg:aug:feedback}). This LLM call returns  NL instructions on how to correct a specific CTE, which is appended to the agent's context (Line~\ref{alg:augment:context}) to fix each CTE. This process repeats until the augmentation mechanism has provided feedback for every CTE within the agent-generated SQL and each CTE has been fixed. 

Given the SQL draft $s$, \sys retrieves knowledge for $s$ and converts it into feedback $f$, which is appended to the agent's context (Lines~\ref{alg:aug:M_begin}–\ref{alg:aug:M_end}). 
A key observation is that different sub-queries within a single SQL query may contain different misconceptions, and therefore correcting them requires different knowledge statements. To improve precision in retrieval and knowledge application, we retrieve knowledge and provide feedback one CTE at a time (the agent is instructed in its initial context to decompose its SQL into as many CTEs as possible). For simple translations, or when the agent produces a SQL with no CTEs, we treat the entire SQL as a single CTE. For each CTE, we first retrieve relevant knowledge (Line~\ref{alg:aug:retrieve}) and then use an LLM to summarize it into actionable feedback for CTE correction (Line~\ref{alg:aug:feedback}). Specifically, the LLM call returns NL instructions for correcting the CTE, which are appended to the agent's context (Line~\ref{alg:augment:context}). This process repeats until feedback has been provided for every CTE and the full SQL has been corrected.

%To address this, refinement is performed at the granularity of individual common table expressions (CTEs).

%In the first step of our \emph{augmentation mechanism}, knowledge is retrieved from the tribal knowledge store $TK$ and converted into actionable feedback on the current SQL draft $s$ using an LLM. 

\textbf{Example}. Fig.~\ref{fig:cte} shows the augmentation process, where first the agent generates a SQL containing CTEs. \sys augments the agent by providing feedback on each CTE, one by one. %\texttt{polo\_items}, contains the products in the \texttt{purchase} table that are from the \emph{Polo} brand. This CTE 
For the first CTE, \sys retrieves relevant knowledge from TK-Store, stating that the \texttt{name} column should be used instead of \texttt{brand}. \sys then uses this knowledge to provide feedback to the NL2SQL agent about how to modify the CTE. The NL2SQL agent then applies the feedback, fixing the first CTE while the second CTE remains incorrect. \sys then moves on to the second CTE, now retrieving different knowledge from TK-Store, and producing feedback to modify how to filter the date column in the CTE. The NL2SQL agent applies this feedback and produces a correct CTE. 

\section{Evaluation}
\label{sec:evaluation}

We evaluate \sys as an augmentation mechanism for NL2SQL agents. In Sec~\ref{sec:evalstup}, we discuss the evaluation setup. End-to-end comparison with baselines on execution accuracy and latency are summarized in Sec~\ref{sec:end2end}, and we additionally demonstrate \sys can augment various NL2SQL agents in Sec~\ref{sec:agentar}. Furthermore, we conduct ablation studies on knowledge content and retrieval (Sec~\ref{sec:ablation-content-vs-usage}), and provide sensitivity analysis (Sec~\ref{sec:sensitivity}) and error analysis (Sec~\ref{sec:error_analysis}).
%Across all methods, the underlying NL2SQL agent~$A$ (Section~\ref{sec:definition}) is held fixed. Methods differ only in the augmentation mechanism~$M(\mathcal{E}, C_t)$, i.e., how experience from the training set is discovered, stored, retrieved, and applied as feedback in the agent’s context during SQL generation and/or refinement steps.

\subsection{Setup}
\label{sec:evalstup}

\subsubsection{Tasks and Datasets.}
We evaluate \sys on two standard \\ 
NL2SQL benchmarks, Spider~2~\cite{lei2024spider} and BIRD-mini-dev~\cite{li2024can}.
Spider~2 consists of three different subsets, each for a different database system (and dialects), specifically, SQLite, BigQuery, and Snowflake. We refer to these three subsets of Spider~2 as SQLite, BQ, and SF, respectively. %which differ in engine, schema, and SQL dialect, ensuring results are not driven by artifacts of any single backend.
We use a filtered BIRD subset that removes known annotation errors~\cite{jin2026pervasive}.
% This evaluation set up ensures that improvements from \sys are not attributable to backend-specific behavior or workload idiosyncrasies. 

Table~\ref{tab:benchmarks} reports the data splits used in our experiments.
For each dataset, we use a randomly sampled 25\% subset of queries to construct the experience set and evaluate on the remaining 75\%.
Experience sets are constructed independently for each dataset without experience or knowledge being shared. % across SQLite, BigQuery, or Snowflake.
All experience and knowledge are derived exclusively from the training queries, and results are reported on held-out test queries.
As agent NL2SQL generation is subject to variance across runs, we evaluate all metrics as an average over two independent runs.

\begin{table}[t]
\vspace{-0.5em}
  \centering
  \small
  \caption{Evaluation benchmarks and data splits. SQLite, BigQuery, and Snowflake are 3 datasets within Spider~2.}
  \label{tab:benchmarks}
  \setlength{\tabcolsep}{6pt}
  \begin{tabular*}{\columnwidth}{l@{\extracolsep{\fill}}ccc}
    \toprule
    Datasets & Full & Train & Test \\
    \midrule
    Spider~2 (SQLite)     & 135 & 34 & 101 \\
    Spider~2 (BigQuery)  & 205 & 50 & 155 \\
    Spider~2 (Snowflake) & 208 & 29 & 179 \\
    BIRD                 & 283 & 72 & 211 \\
    \bottomrule
  \end{tabular*}
\end{table}

% \end{itemize}

\begin{figure*}[t!]
    \centering
    \vspace{-0.5em}
    \includegraphics[scale=0.445]{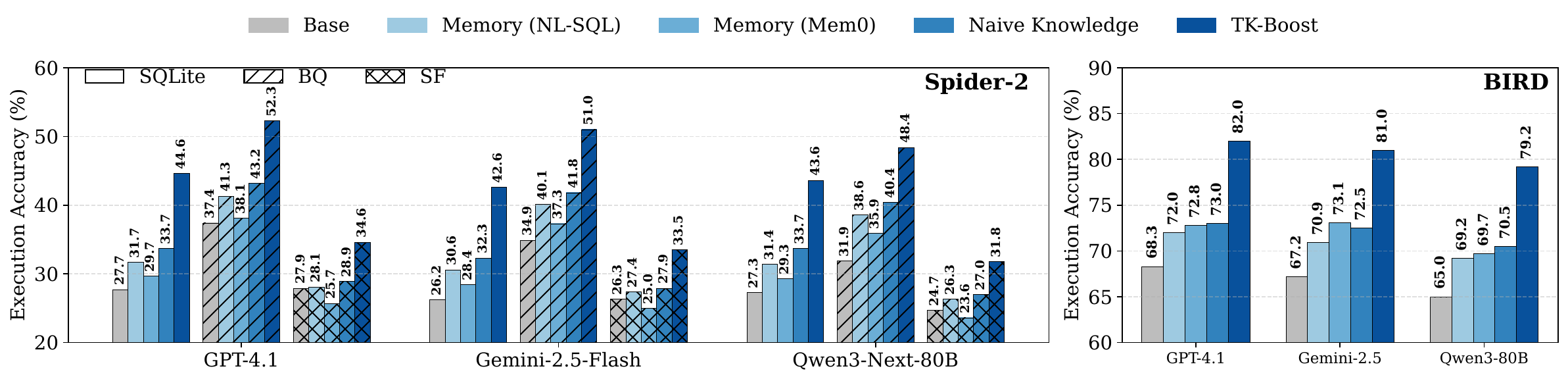}
    \caption{Execution accuracy across NL2SQL methods. 
    (Left) Spider~2 results across SQLite, BigQuery, and Snowflake datasets. (Right) BIRD results. TK-Boost consistently outperforms prior memory and knowledge-based baselines across models.}
    \label{fig:main-results-plots}
\end{figure*}

\vspace{-0.2cm}
\subsubsection{Baselines and Augmentation Mechanisms.}
For the set of experiments,  we fix an NL2SQL agent~$\mathcal{A}$. We compare \sys with different augmentation mechanisms that use the experience set to improve the accuracy of $\mathcal{A}$; thus, each of our baselines has an augmentation mechanism that only modifies the \emph{content} of the agent’s context. We describe the agent $\mathcal{A}$, and the models used in Sec.~\ref{sec:exp:models_agents}.
%Given a natural-language query, the agent iteratively generates and executes SQL queries and refines its output until a final query is produced.
%All methods are provided an uncapped inference budget: the maximum number of agent turns, and the number of retrieved items during top-K similarity search are not modified between baselines.

% \sep{how do we set $k$? What value is it in our baselines?}. \sep{I'd add a section called parameter settings and just discuss all these different parameters, e.g., maximum number of turns, number of retrieved items, etc. }

\begin{itemize}

\item \textbf{Base NL2SQL Agent.}
The unaugmented agent~$A$, corresponding to  
$M(\mathcal{E}, C_t) = C_t$.

\item \textbf{Memory Augmentation.}
The memory-augmentation mechanism described in Section~\ref{sec:memory}. 
% \sep{whta's the difference between these two? Which one is the one in Sec. 3.1?}
% \begin{itemize}
%     \item \textbf{NL--SQL pairs.}
%     NL--SQL pairs consisting of only the natural language question and the corresponding ground truth SQL are retrieved via natural-language similarity from the experince set. Relevant pairs are injected into the initial context~$C_0$ as in-context examples.
    
%     \item \textbf{Mem0.}
%     The Mem0 execution-trace memory mechanism (discussed in Sec~\ref{sec:memory}) stores full agent execution \emph{traces} from training queries, including all reasoning from the agent as it attempts a question.  
%     At test time, the retrieved traces are injected into the initial context~$C_0$ based on NL similarity. \alvin{how is similarity defined}
% \end{itemize}

\begin{itemize}
    \item \textbf{Mem0.}
    Mem0~\cite{mem0} is used to augment the initial context~$C_0$. It store agent execution traces from training queries. At test time, traces are retrieved using a cosine-similarity embedding search over NL queries.

    \item \textbf{NL--SQL Pairs.}
    Instead of the entire execution trace as in Mem0, here only the NL question and its ground-truth SQL are provided. NL--SQL pairs are retrieved from the experience set using cosine similarity over NL-question embeddings.

\end{itemize}

\item \textbf{Naive Knowledge.}
An NL2SQL agent augmented with natural-language knowledge generated from NL--SQL pairs in the experience set with a single LLM call. Retrieved knowledge is injected into the initial context~$C_0$ based on natural-language similarity, without explicit applicability conditions.

\item \textbf{\sys (Ours).} This is our framework, as presented in Alg.~\ref{alg:knowledge_application}.
%Our system converts logical difference sets extracted from SQL edits between agent SQL and ground-truth SQL in the experience set into structured knowledge with explicit applicability conditions over SQL operators and schema elements.  
%At test time, relevant knowledge is retrieved and applied to each CTE in the agent’s initial SQL attempt by augmenting the intermediate context~$C_t$ across a series of refinement iterations.

\end{itemize}

\subsubsection{Models and Agents}\label{sec:exp:models_agents} \hfill \\
\noindent\textbf{Embeddings.} We perform semantic similarity search using OpenAI \texttt{text-embedding-3-large} as the embedding model. \\
\noindent\textbf{Models.} Our evaluation uses four different LLMs: OpenAI GPT-4.1, Gemini-2.5-Flash, and Qwen3-Next-80B as well as Agentar-Scale-SQL-32B~\cite{wang2025agentar}. The first three span common pre-trained models used for NL2SQL~\cite{lei2024spider, liu2024survey}, and the fourth is a BIRD-specialized finetuned model---we use this model to show \sys can be used to augment agents even when they use models fine-tuned for a specific dataset. We use provider APIs for the closed-source GPT and Gemini models, and host the open-source Qwen model and Agentar-Scale-SQL-32B on two 80GB H100 GPUs. %To evaluate \sys as a bolt-on augmentation for existing NL2SQL systems (Sec.~\ref{sec:agentar}), ,  \\

% \noindent\textbf{Agents.} Unless otherwise stated, our NL2SQL agent (i.e., the function $\mathcal{A}$ Sec.~\ref{sec:definition}) is a ReAct agent~\cite{yao2023react} powered by one of the models above, which is given a tool to execute SQL queries on the DB. Each turn of the ReAct agent contains a reasoning and action component, and the SQL query execution is the agent's action. % and iteratively generates SQL to answer a query, as discussed in Sec.~\ref{sec:definition}. 
% To evaluate generality of \sys across various agentic architectures, we also evaluate \sys when using the ReFORCE~\cite{deng2025reforce} NL2sQL agent, which uses a more complex architecture, with multiple modules such as schema compression, parallel candidate generation, and majority voting.  %, .

\noindent\textbf{NL2SQL Execution Frameworks.}
Unless otherwise stated, the NL2SQL function $\mathcal{A}$ (Sec.~\ref{sec:definition}) follows a ReAct-style execution framework~\cite{yao2023react}, where a single model iteratively generates SQL queries, executes them against the database, and refines them based on execution feedback. To evaluate the generality of \sys, we also instantiate $\mathcal{A}$ using ReFORCE~\cite{deng2025reforce}, which replaces the single-model execution loop with a multi-component pipeline that includes schema compression, parallel SQL candidate generation, and majority voting, and uses multiple LLM calls.

\subsubsection{Metrics.}
We report the following metrics:

\begin{itemize}
    \item \textbf{Execution Accuracy.}
    Fraction of queries whose final SQL executes successfully and returns the correct result.

    \item \textbf{Robustness.}
    Fraction of queries answered correctly by the base NL2SQL agent but incorrectly after augmentation, measuring the risk of knowledge misapplication.

    \item \textbf{Latency and Turns.} Average end-to-end latency, and agent interaction rounds needed to produce the final SQL query, characterizing inference-time overhead.
\end{itemize}

\subsubsection{Parameter Settings.}
\label{sec:param}
Across all baselines, we use an uncapped inference budget with identical limits on the maximum number of agent turns. We retrieve top-$K{=}5$ items for approaches that use semantic similarity for retrieval, with $K{=}5$ chosen as it performed the best across baselines. All LLM calls use temperature $t{=}0.5$ and allow up to 128k tokens per run.
% During knowledge discovery (Alg.~\ref{alg:knowledge-generation}), Jaccard-similarity guardrail uses a threshold of $\alpha{=}0.3$.

% % --------------------------------------------------------------------------

% \textbf{\textcolor{red}{NEW OUTLINE:}}

% \textcolor{red}{ Sec 6.2}
% Para 1 - We are great. BOLD.
% 1 sentence can be much briefer. Just say retrieves bad examples, and gets worse with trace for mem0. No need for examples or "details".

% Tell the main trend: base is worse; memory helps a little (x\%); this is because it uses some of the \textbf{very similar} gold, and for any differences, it retrieves just bad queries and NO TRANSFER.

% for knowledge, cite the ACL, say as it is - can we?

% row 5 is an ablation - our --

% 6.2 should be small - one to highlight we are good and the trend, one on why others are bad

% Talk on all: accuracy, robustness, and turns/latency.

% \textcolor{red}{Sec 6.3}
% Tell that in table 1 we show for 3 MODELS ALREADY, and then show why, and that we show for SOTA SOTA also.
% Talk in the main results only for the applicability of augmenting to SOTA.

% \textcolor{red}{Sec 6.4}
% Ablations of components

% \textcolor{red}{Sec 6.5}
% Sensitivity analysis

% \textcolor{red}{Sec 6.6}
% Error analysis (don't forward reference the table) FOLD IN with case study
% -- make the error categorisation explicit \textbf{here}

\subsection{End-to-End Results}
\label{sec:end2end}

% \joey{I rewrote the first sentence (started at a modified version of the second sentence) to emphasize the gains. }
\subsubsection{Execution Accuracy}
% Fig.~\ref{fig:main-results-plots} reports execution accuracy across all datasets on multiple base models. 
Across models and datasets (see Fig.~\ref{fig:main-results-plots}), \textbf{ \sys yields accuracy gains of up to 16.9\% on Spider-2 and 13.7\% on BIRD}, representing a substantial improvement over the Base Agent.
% \sep{Use capitalized if we are referring to the baseline. Might even be better if we give the baseline and actual name}. 
These gains significantly exceed the modest improvements from memory baselines (3.9\%, 4.5\% respectively) and naive knowledge baselines (5.8\%, 4.7\% respectively), and are significant strides in execution accuracy on these benchmarks, where gains of even 5\% have historically required model finetuning or complex multi-agent pipelines with significant budget.
In contrast, across all models, the Base NL2SQL Agent performs poorly with an execution accuracy of 24.7\%--37.4\%  on Spider~2 and 65.0\%--68.3\% on BIRD, highlighting the difficulty of enterprise-scale text-to-SQL. 
% \joey{Can we say something about how these improvements compare to other improvements in prior work.  For example, These improvements exceed improvements made by other recent innovations in models and agent loops.}

% \joey{I separated this out into a new paragraph and moved the base agent results to the previous paragraph.}
Memory-based augmentation yields only small gains (4.5\%), primarily in cases where test questions closely match prior examples, particularly on datasets with many similar queries on a database (e.g., Spider~2 BigQuery). For most questions, memory retrieval surfaces irrelevant or misleading prior experiences, limiting generalization. The Naive Knowledge baseline shows modest gains by compressing past experience into generalized knowledge statements. However, this knowledge remains brittle, tied to surface-level query differences, and retrieval failures continue to occur.

\sys addresses these 
% \joey{what limitations?  Restate.}
limitations in knowledge resuability by extracting tribal knowledge from past errors and defining structured applicability conditions for usage. As a result, it consistently improves across performance models, even when the underlying agent misconceptions differ, highlighting that structured knowledge is more effective than raw memory or surface-level knowledge generation. We provide an example and compare retrieved knowledge and its impact on the agent in Sec~\ref{sec:error_analysis}.
\vspace{-5pt}

\begin{figure}[t]
    \centering
    \includegraphics[width=0.75\columnwidth]{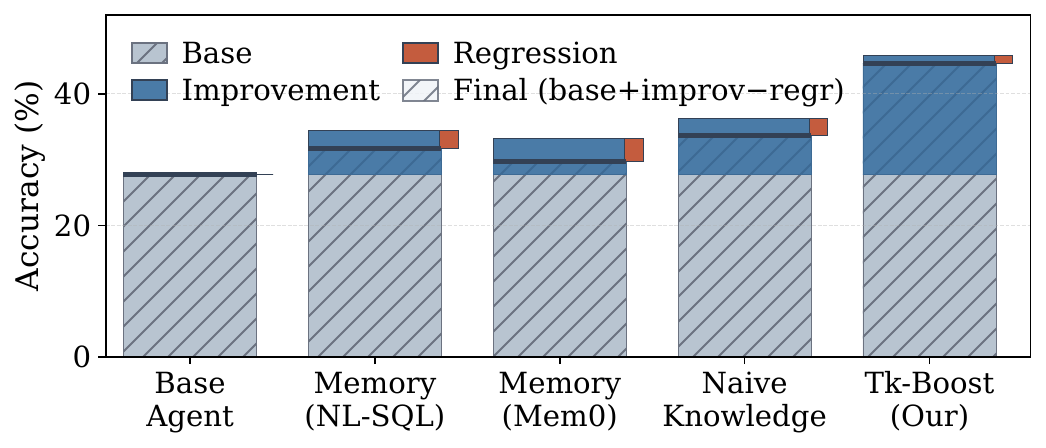}
      \caption{Accuracy decomposition on Spider~2 SQLite (GPT-4.1).
  Improvement: fraction of Base Agent's errors now fixed.
  Regression: fraction of Base Agent's correct predictions now broken.
  Net = Improvement $-$ Regression.}
    \label{fig:regression}
\end{figure}

\subsubsection{Robustness}
We also analyze how robust each method is by decomposing its outcomes into
\emph{improvement} (the fraction of Base Agent errors that are corrected), and
\emph{regression} (the fraction of Base Agent correct predictions that become incorrect) after augmentation. Fig.~\ref{fig:regression} illustrates this on Spider~2 SQLite (GPT-4.1).

While memory-based methods yield $>$5\% accuracy improvements before regressions, the net improvement over the Base Agent is limited (2.0\% to 4.0\%) as there are regressions of about 3\% of initially correct questions. 
The high-recall nature of memory-based methods can mislead the agent by introducing irrelevant information from past experiences as purely semantic similarity is used for retrieval, irrespective of context. Using Naive knowledge achieves even larger raw improvements (8.6\%), but regressions are still present (2.6\%). Utilizing a singular LLM call to generate knowledge from an example NL--SQL pair introduces hallucination risk, as the LLM is not given context to understand why a mistake was made, and as a result, the knowledge misleads the agent when applied. We detail an example of this in Sec~\ref{sec:error_analysis}.

% \alvin{did this actually happen? if so say that's what we observed as the reason for regression} \asim{add short example here, can be the same as 6.2.1 example :
% For example, for rolling-window and \texttt{LAG} queries (described in detail in Sec.~\ref{sec:error_analysis}), 
% we observe regressions because naive knowledge states to use window functions but omits the required ordering:
% window functions must be computed over the full history before date filtering.
% }

\sys achieves both the greatest improvement and the least regression. It corrects 18.1\% of the Base Agent’s errors while only 1.2\% of correct questions regress,
yielding a net gain of +16.9\%. Structured retrieval reduces errors occurring from retrieving irrelevant information, and as a result, the robustness is highest, with minimal regressions occurring due to ambiguous queries. 

\vspace{-0.1cm}
\subsubsection{Latency and Number of LLM Calls}

Table~\ref{tab:efficiency} reports end-to-end latency and total number of LLM calls performed per query. Note that the function $\mathcal{A}$ is a  ReAct agent that makes one LLM call per iteration; \sys makes additional LLM calls for retrieval and feedback generation. Queries contain 2.2 $\pm$ 2.3 CTEs on average.

\sys increases latency by 7.4 seconds over the base agent (29.6$\rightarrow$37.0s) while improving execution accuracy by 16.9 percentage points. Total LLM calls rise modestly from 12.0 to 15.9 per query (+3.9), driven by CTE-level feedback refinement that adds targeted refinement steps to correct logical errors. Overall, modest additional inference yields substantial accuracy gains.

\begin{table}[t]
  \centering
  \footnotesize
  \caption{Total LLM calls and end-to-end latency per query on Spider~2 SQLite (GPT-4.1). 
  \sys incurs modest overhead while delivering substantially higher execution accuracy.}
  \label{tab:efficiency}
  \setlength{\tabcolsep}{6pt}
  \begin{tabular}{lcc}
    \toprule
    \textbf{Method} & \textbf{\# LLM Calls} & \textbf{Latency (s)} \\
    \midrule
    Base NL2SQL Agent      & 12.0 $\pm$ 2.8 & 29.6 $\pm$ 9.2 \\
    Memory (NL--SQL Pairs) & 13.6 $\pm$ 3.3 & 32.4 $\pm$ 10.1 \\
    Memory (Mem0)         & 14.5 $\pm$ 3.9 & 34.4 $\pm$ 10.8 \\
    Naive Knowledge       & 13.9 $\pm$ 3.6 & 33.2 $\pm$ 10.4 \\
    {\sys (Ours)}         & 15.9 $\pm$ 4.5 & 37.0 $\pm$ 13.3 \\
    \bottomrule
  \end{tabular}
\end{table}

\vspace{-0.3cm}

\subsection{Applicability to Other NL2SQL Agents}
\label{sec:agentar}

\begin{figure}[t!]
    \centering
    \begin{minipage}{0.48\columnwidth}
        \centering
        \includegraphics[width=\linewidth]{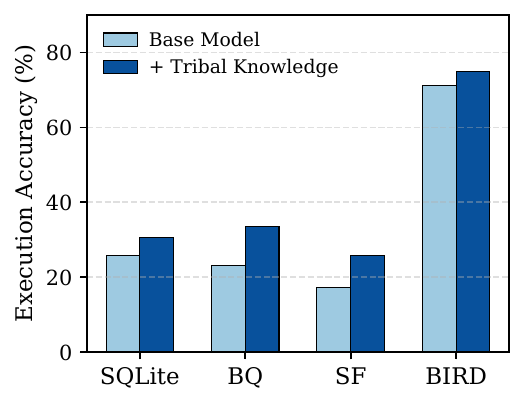}
        \vspace{2pt}
        \small (a) Agentar-Scale-SQL (32B)
    \end{minipage}
    \hfill
    \begin{minipage}{0.48\columnwidth}
        \centering
        \includegraphics[width=\linewidth]{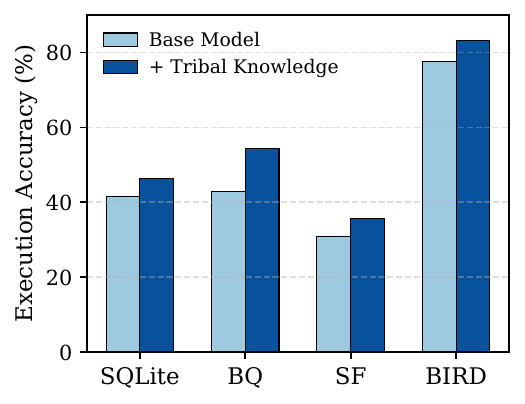}
        \vspace{2pt}
        \small (b) ReFoRCE (GPT-4.1)
    \end{minipage}

    \caption{Execution accuracy with and without \sys augmentation for existing NL2SQL solutions. \sys successfully increases execution accuracy on both Spider~2 and BIRD on Agentar-Scale-SQL and ReFoRCE by utilizing tribal knowledge for SQL refinement.}
    \label{fig:sota_with_tk}
\end{figure}

\sys is a bolt-on solution that can be applied to any NL2SQL agent. % for various improving existing NL2SQL systems. 
In Sec.~\ref{sec:end2end}, we discussed its applicability to an agent following a basic ReAct framework and with different base models, and here we show it applies to specialized finetuned models and complex agent architectures that many of the best performing benchmark solutions utilize.
We select solutions, namely Agentar-Scale-SQL~\cite{wang2025agentar} and ReFORCE~\cite{deng2025reforce} that achieve the highest execution accuracy on Spider~2 and BIRD, and have code available. Fig.~\ref{fig:sota_with_tk} reports execution accuracy for each solution, where \sys is applied only at refinement time to improve generated SQL and does not interfere with the agent's SQL generation.

Although Agentar-Scale-SQL-32B performs strongly on BIRD (71.2\%) as it is finetuned for it, it generalizes poorly to Spider~2 datasets. Applying \sys yields consistent gains across all datasets, improving accuracy on SQLite (25.3\%$\rightarrow$30.8\%), BigQuery (23.4\%$\rightarrow$33.6\%), Snowflake (17.2\%$\rightarrow$25.6\%), and BIRD (71.2\%$\rightarrow$74.8\%). These improvements indicate that \sys can identify and correct misconceptions even for an NL2SQL model that was already finetuned for the BIRD dataset ~\cite{li2024can}. 
% \sep{should emphasize a bit more that you're actually improving the model that was already fine-tuned for BIRD}
\sys also improves ReFORCE, a strong multi-step agentic pipeline, raising execution accuracy from 41.5\%$\rightarrow$46.3\% on SQLite, 43.0\%$\rightarrow$54.4\% on BigQuery, and 31.0\%$\rightarrow$35.7\% on Snowflake, and 77.5\%$\rightarrow$83.1\% on BIRD. 

Overall, \sys functions as a model and \emph{architecture-agnostic bolt-on} by improving both specialized single-model systems and complex multi-step pipelines.

\begin{table}[t]
\centering
\footnotesize
\caption{\textbf{Content ablation.}
Retrieval is done by matching query features for each CTE; the application is for refining each CTE. Differences reflect knowledge content only.}

\setlength{\tabcolsep}{6pt}
\begin{tabular}{lc}
\toprule
\textbf{Knowledge Source (content only)} & \textbf{Accuracy (\%)} \\
\midrule
No knowledge (Base Agent) & 27.7 \\
Memory (NL--SQL Pairs) & 32.7 \\
Naive Knowledge & 36.7 \\
Non-Atomic Correction Statements & 39.8 \\
Correction Statements (Non-generalized) & 41.6 \\
% Correction-Based Knowledge & 39.7 \\
\textbf{Tribal knowledge (\sys)} & \textbf{44.6} \\
\bottomrule
\end{tabular}

\label{tab:ablation-content}
\end{table}

% \vspace{-0.5cm}
\subsection{Ablation Studies: Knowledge Type vs.\ Knowledge Retrival and Application}
\label{sec:ablation-content-vs-usage}

We ablate tribal knowledge augmentation along two dimensions:
(i) the \emph{content} of the knowledge statements, and
(ii) the \emph{retrieval and application mechanism}.
All experiments are conducted on Spider~2 SQLite using the same base NL2SQL agent.

% ---------------------------------------------------------------------------
% \subsubsection{Effect of Knowledge Type (Content)}
% \label{sec:ablation-content}

% Table~\ref{tab:ablation-content} isolates the effect of knowledge \emph{content}.
% All rows use an identical usage pipeline:
% knowledge is retrieved using query characteristics identified in each CTE and applied to correct the SQL at a
% CTE-level. For rows using a knowledge source, the content of the knowledge is varied between memory-based knowledge, naive knowledge, an unprocessed logical error set, and tribal knowledge.

% Under the same retrieval and application mechanism, replacing naive knowledge with correction-based knowledge increases execution accuracy from
% 32.7\% to 39.7\%.
% Using tribal knowledge further
% increases accuracy to 44.7\%.
\vspace{-1.5mm}
\subsubsection{Knowledge Type}
\label{sec:ablation_content}

Table~\ref{tab:ablation-content} isolates the effect of knowledge content while keeping retrieval and CTE-level refinement fixed.

Both NL--SQL pair memory and naive knowledge improve execution accuracy (27.7\%→32.7\%, +5.0\%; 32.7\%→36.7\%, +4.0\%) yet fail to resolve logical errors, as they do not encode the underlying corrections. Correction statements, before generalizing into knowledge with applicability conditions, raise accuracy to 41.6\% by capturing the required logical fix, but their sample-specific scope restricts reuse. Non-atomic corrections, unconstrained to a single clause-level edit, perform worse with an accuracy of 39.8\%, as these corrections are broad rewrites that approximate the ground-truth SQL rather than capturing the agent's misconception.

Tribal knowledge achieves the highest accuracy at 44.6\% by generalizing corrections and associating them with applicability conditions, enabling reusable fixes that encode both the required logical change and the contexts in which it should be applied.

% ---------------------------------------------------------------------------
% \subsubsection{Effect of Retrieval and Application (Usage)}
% \label{sec:ablation-usage}

% Table~\ref{tab:ablation-usage} demonstrates differences in performance for retrieval strategies over the same set of tribal knowledge statements.

% In the first row, tribal knowledge is retrieved once using NL-query similarity
% and injected into the initial prompt, with no post-generation refinement.
% Structured retrieval using \textsc{TK-Retrieve} is not applicable in this setting because no initial SQL draft is
% available.

% In the second row, an initial SQL is generated without knowledge.
% Knowledge is then retrieved based on either the full SQL structure or
% for each CTE, merged, and used to refine
% the full SQL. Retrieving knowledge individually for each CTE yields the same rule set after merging,
% resulting in identical accuracy (39.7\%).

% In the third row, SQL is decomposed into CTEs. Knowledge is retrieved separately for each CTE based on its characteristics
% and used to refine that CTE only.
% This configuration achieves the highest execution accuracy (44.7\%).

\subsubsection{Knowledge Retrieval and Application}
\label{sec:ablation-usage}

Table~\ref{tab:ablation-usage} isolates the effect of retrieval and application while keeping knowledge content fixed to tribal knowledge statements.

Retrieving knowledge using NL-query similarity and injecting it before SQL generation achieves 31.7\% accuracy. Applying the same knowledge during refinement on either the full SQL or on each CTE improves accuracy to 35.7\% (+4.0\%). This approach remains constrained by coarse retrieval that often doesn't retrieve knowledge that actually addresses the error.

Retrieving knowledge from the generated SQL by matching query features increases accuracy to 39.7\% (+4.0\%). Applying this knowledge for refinement on either the full SQL or each CTE yields similar accuracy (39.7\% vs.\ 41.6\%), indicating that gains at this stage arise primarily from improved retrieval.

Retrieving and applying knowledge independently for each CTE achieves the highest accuracy at 44.6\% (+4.9\%), showing that knowledge is most effective when retrieval and application are used to identify and fix errors within each SQL sub-task at a finer granularity, rather than using coarse-grained feedback for broad suggestions on the entire SQL query. 
% \alvin{by not together you mean retrieve and apply separately? how can that work? you meant fix the entire query?}

% \vspace{-1em}
\subsection{Sensitivity}
\label{sec:sensitivity}

We analyze the sensitivity of \sys to (i) the amount of knowledge retrieved and applied at refinement time and (ii) the size of the experience set used for knowledge discovery. All experiments use GPT-4.1 and are evaluated on Spider~2 SQLite and Snowflake, with all other components held fixed.

\begin{table}[t]
\centering
\footnotesize
\caption{\textbf{Usage ablation.}
Knowledge content is fixed to tribal knowledge statements.
Cells marked ``--'' correspond to configurations that are not applicable.}
\setlength{\tabcolsep}{5pt}
\begin{tabular}{lccc}
\toprule
& \multicolumn{3}{c}{\textbf{Retrieval Strategy}} \\
\cmidrule(lr){2-4}
\textbf{Application Method} &
\textbf{NL query} &
\textbf{SQL} &
\textbf{CTE} \\
\midrule
Base Agent + NL-query-Level & 31.7 & -- & -- \\
Base Agent +  SQL-Level & 35.7 & 39.7 & 39.7 \\
\textbf{Base Agent + CTE-Level} & 35.7 & 41.6 & \textbf{44.6} \\
\bottomrule
\end{tabular}

\label{tab:ablation-usage}
\end{table}

\begin{table}[t]
\centering
\footnotesize

\caption{Sensitivity analysis for Tribal Knowledge (T3).
Execution accuracy (\%) using GPT-4.1 on Spider~2 SQLite and Snowflake.
All other components are held fixed.}
\begin{minipage}[t]{0.48\linewidth}
\centering
\begin{tabular}{lccc}
\toprule
\textbf{} & Zero & All & Our \\
\midrule
SQLite      & 27.7 & 35.6 & \textbf{44.6} \\
Snowflake  & 27.8 & 30.2 & \textbf{34.6} \\
\bottomrule
\end{tabular}

\vspace{2pt}
\textbf{(a) \# knowledge statements applied.}
\end{minipage}
\hfill
\begin{minipage}[t]{0.48\linewidth}
\centering
\begin{tabular}{lccc}
\toprule
\textbf{} & $10\%$ & $25\%$ & $50\%$ \\
\midrule
SQLite      & 29.7 & 44.6 & \textbf{51.5} \\
Snowflake  & 27.4 & 34.6 & \textbf{39.7}\\
\bottomrule
\end{tabular}

\vspace{2pt}
\textbf{(b) Experience set size.}
\end{minipage}

\label{tab:sensitivity}
\end{table}

% \vspace{-1em}

\subsubsection{Number of Knowledge Statements}

Table~\ref{tab:sensitivity}(a) compares three settings: no knowledge, providing the full contents of the knowledge store, and retrieval using \sys's retrieval function.

On SQLite, execution accuracy improves from 27.7 to 35.6 when all knowledge is applied (+7.9\%), and further to 44.6 with scoped retrieval (+9.0\% over all knowledge). On Snowflake, accuracy increases from 27.8 to 30.2 (+2.4\%) with all knowledge and to 34.6 (+4.4\%) with scoped retrieval.

The all-knowledge setting exposes the model to the entire knowledge store extracted from the 25\% training split (80 knowledge statements for SQLite and 104 for Snowflake). In contrast, scoped retrieval with \sys selects knowledge per CTE, typically retrieving 2--8 statements conditioned on the query context (average of 4.2). Across both backends, scoped retrieval consistently outperforms applying all knowledge, indicating that selective, context-aware retrieval is critical for effective knowledge application. 
% \alvin{so what's the takeaway? scoped is the best? \asim{yes this was the intended takeaway. if that's not clear in text now we can make that more explicit?}}

% \sep{might be good provide actual statistics about (1) total number of knowledge statements and (2) how many we use on average across all datasets in a table} 

\subsubsection{Experience Set Size}

Table~\ref{tab:sensitivity}~(b) reports execution accuracy as a function of the fraction, $p$, of the total number of Spider~2 queries (SQLite and Snowflake) used to construct tribal knowledge. On SQLite, accuracy rises from 29.7 at $p{=}10\%$ to 44.6 at $p{=}25\%$ (+14.9\%), and further to 51.5 at $p{=}50\%$ (+6.9\%). Snowflake shows a similar trend, improving from 27.4 to 34.6 (+7.2\%) and to 39.7 (+5.1\%). Most gains are realized by $p{=}25\%$, with diminishing returns thereafter. Increasing the experience set size results in test-set accuracy increases on both SQLite and Snowflake, as more data misconceptions are captured when more experience becomes available. 

% \begin{figure}[t]
%     \centering
%     % \vspace{-1em}
%     \includegraphics[width=0.8\columnwidth]{figures/error_breakdown_absolute-1.pdf}
%     \caption{Error composition on Spider~2 SQLite (GPT-4.1) for each baseline. \sys significantly reduces data usage errors while also reducing data selection errors.}
%     \label{fig:error-composition}
% \end{figure}

\vspace{-1em}

\begin{figure}[t]
    \centering
    % \vspace{-1em}
    \includegraphics[width=0.8\columnwidth]{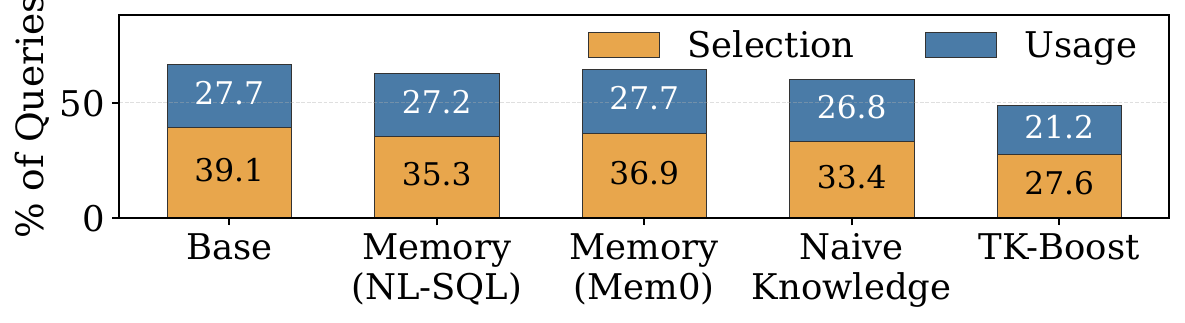}
    \caption{Error composition on Spider~2 SQLite (GPT-4.1) for each baseline. \sys significantly reduces data usage errors while also reducing data selection errors.}
    \label{fig:error-composition}
\end{figure}

\begin{table}[t]
\centering
\footnotesize
\caption{Representative tribal knowledge statements, grouped by misconception type.}
\label{tab:knowledge-samples}
\setlength{\tabcolsep}{4pt}
\renewcommand{\arraystretch}{1.1}
\begin{tabular}{p{0.165\linewidth} p{0.78\linewidth}}
\toprule
\textbf{Type} & \textbf{Tribal Knowledge Statement} \\
\midrule
\multirow{3}{*}{\textbf{Data-Selection}} 
& Filter calendar-year queries using date fields (e.g., \texttt{sales.date}), not fiscal-year attributes. \\

% & Exclude rows with non-numeric volume encodings (e.g., \texttt{12K}, \texttt{3M}, \texttt{-}) when selecting data for numeric analysis. \\

& Use entity-level metrics (e.g., business identifiers) for aggregation not surrogate join keys. \\

& Use \texttt{DISTINCT order\_id} to define the entity when selecting records for order-level metrics. \\

\midrule
\multirow{4}{*}{\textbf{Data-Usage}} 
& Compute window functions over the full date range; apply filters only after window evaluation. \\

& Validate join cardinality before aggregation; unintended many-to-many joins inflate results. \\

& Convert epoch fields to timestamps before applying date predicates. \\

\bottomrule
\end{tabular}
\end{table}

\subsection{Examples of Agent Misconceptions}
\label{sec:error_analysis}

We analyze the percentage of errors that occur due to \emph{data-selection} misconceptions versus \emph{data-usage} misconceptions. After filtering out errors that occur due to agent runtime errors and instances where the agent does not produce a SQL, we utilize a SQL parser to extract tables and columns from each incorrect agent SQL and the corresponding ground truth SQL. If a mismatch is identified, we label it as a data-selection error, and otherwise we label the error as data-usage. We note that in scenarios where data-usage errors occur alongside data-selection errors, we consider data-selection as the main error source. 

Fig.~\ref{fig:error-composition} illustrates the percentage of errors of each type across methods. Memory-based augmentation struggles to resolve any data usage error while providing a modest reduction in data selection errors (39.1\% → 35.3\%). Naive knowledge augmentation reduces data selection errors further (33.4\%) but provides minimal resolution on data usage errors. While prior methods can reduce data-selection errors, data-usage misconceptions remain a substantial issue. Recalling examples directly from the experience set may reveal which tables and columns are appropriate for similar questions on a complex database, but the way to utilize the data is much harder to gather from direct memories or one-shot LLM-generated naive knowledge. In contrast, \sys reduces data-selection errors from 38.0\% to 27.6\% and data-usage errors from 28.0\% to 21.2\%, yielding a 15.9\% improvement in execution accuracy overall. 
% Storing knowledge with structured applicability conditions allows the NL2SQL agent to resolve data usage issues by understanding SQL contexts where it has misconceptions (e.g., performing a filter on date type columns) and how to resolve them.

Table~\ref{tab:knowledge-samples} shows representative tribal knowledge statements learned by \sys that encode reusable corrections for each error type. For example, a Spider 2 SQLite query \texttt{local077} asks for three-month rolling averages and lagged values over 12 months on bank sales data; the agent applies a date filter before computing rolling averages and \texttt{LAG} values. However, this truncates the historical prefix required for window evaluation and produces biased aggregates. NL-similarity memory retrieves a ``monthly revenue average'' query that does not use window functions or \texttt{LAG}, and therefore only shows a plain \texttt{AVG} after filtering. Naive knowledge contains a broad instruction to "use WINDOW to compute window function" but omits the required ordering nuance; window functions must first be computed over the full date range before date filtering--so the same error repeats. \sys instead provides the first data-usage rule in Table~\ref{tab:knowledge-samples}--compute window functions over the full date range and apply filters after--to the agent during refinement of the CTE, where it creates the window, precisely resolving this error while keeping other parts of the query untouched.

\section{Related Work}
\label{sec:related}
\noindent\textbf{NL2SQL Systems with Agents.}
Many recent NL2SQL systems adopt agentic designs that decompose SQL generation into iterative planning, tool use, and verification \cite{lei2024spider, biswal2024text2sql, li2024codes, schmidt2025sqlstorm, chen2025reliable, rahaman2024evaluating, xiong2025multi, pourreza2024chase, liu2025supporting, shen2025magesql, 10.14778/3681954.3681960}. Some recent representative approaches are ReFORCE that introduces a multi-step agent architecture where candidate SQLs are generated, voted on, and refined for enterprise-scale databases \cite{deng2025reforce}, and AgenticData that augments agents with schema exploration, table sampling, and intermediate result inspection tools \cite{sun2025agenticdata}. %Unlike execution-feedback or self-reflection methods that repair only the current query \cite{shen2025magesql}, \sys extracts persistent corrective knowledge that generalizes across future queries.

\sys operates orthogonally to these frameworks. Rather than modifying agent planning or tools, it acts as a bolt-on layer that provides feedback to the agents by discovering and using tribal knowledge. Given a draft SQL, \sys identifies recurring logical misconceptions and provides targeted feedback on subtasks. %, improving NL2SQL systems with varying agentic architectures . % while preserving the agent’s original reasoning pipeline. 
Our experiment shows that \sys improves  accuracy of NL2SQL agents with varying modeling choices and agentic architectures.  

\vspace{.5em}

\noindent\textbf{Agentic memory.}
Prior work extends LLM agents with persistent memory to overcome context limits, including OS-style memory abstractions (e.g., MemGPT) \cite{li2025memos} and approaches that extract salient information from interaction traces into persistent stores \cite{xu2025mem, chhikara2025mem0, zhong2024memorybank}. Simply retrieving past traces or summaries often reintroduces the same misconceptions, as they do not encode corrective knowledge about how to query the database \cite{biswal2026agentsm, xie2025opensearch}. \sys instead distills corrective, reusable knowledge from past mistakes, which, significantly outperforms memory-based baselines.

\vspace{.5em}

\noindent\textbf{Knowledge base generation for Text-to-SQL.}
Several NL2SQL systems incorporate external knowledge beyond the question and schema \cite{hong2024knowledge, baek2025knowledge, gao2023text, shen2025magesql, zhu2024towards}. Knowledge-to-SQL generates database facts such as column semantics and value conventions to support SQL generation \cite{baek2025knowledge}, but factual knowledge alone does not resolve logical misconceptions caused by data idiosyncrasies ~\cite{huang2024data}. KAT-SQL builds an evolving NL2SQL knowledge base from training queries and schemas, retrieving top-$k$ entries at query time \cite{baek2025knowledge, zhang2023refsql}, similar to the Naive Knowledge baseline in our experiments. This knowledge is driven by surface-level patterns, captures syntactic rather than logical differences, and remains sensitive to semantic retrieval failures, even with ground-truth SQL supervision \cite{baek2025knowledge}. \sys departs from database facts and rule-based or constraint-driven SQL correction \cite{10.14778/3734839.3734847} by grounding knowledge in execution errors that encode how the database should be queried.

\section{Conclusion}

We introduced \sys, a bolt-on framework that augments NL2SQL agents with tribal knowledge distilled from execution mistakes. By storing knowledge with applicability conditions in a structured store and applying knowledge through fine-grained SQL refinement, \sys enables agents to learn how to use databases through experience. Experiments on Spider 2 and BIRD show up to 16.9\% improvement in execution accuracy over baselines across models. \sys is robust and generalizes across agent architectures. Our findings suggest that structured corrective knowledge is necessary for improving data agents and advancing the path toward reliable natural-language access to enterprise databases.
% \input{sections/edits_details}

%\clearpage

\bibliographystyle{ACM-Reference-Format}
\bibliography{sample}

\appendix

\end{document}